\documentclass[
,secnumarabic, tightenlines,amssymb, amsmath,nobibnotes,nofootinbib,aps,
]{revtex4-1}
\usepackage{graphicx}
\usepackage{tabularx}
\usepackage{amsfonts}
\usepackage{amssymb}
\usepackage{bm}%
\usepackage{hyperref}
\usepackage[all]{hypcap}
\usepackage{subfigure}

\usepackage{enumerate}
\expandafter\ifx\csname package@font\endcsname\relax\else
\expandafter\expandafter
\expandafter\usepackage
\expandafter\expandafter
\expandafter{\csname package@font\endcsname}%
\fi

\newcommand{\tr }{\;\mbox{tr}\;}
\newcommand\pfaff{\mbox{Pfaff}\,}

\def\arccosh{\mathop{\rm arccosh}\nolimits}
\def\innprod#1#2{\left\langle#1,#2\right\rangle}
\def\mod{{\rm \,mod\,}}
\def\example#1{
\par\vskip 0.55\baselineskip  minus 0.55\baselineskip\noindent 
{\bf Example}{ \it #1}
\par\penalty 10000\vskip 0.4\baselineskip  minus 0.4\baselineskip\noindent}

\def\se{\mathop\sim\limits^s}
\def\vbundm{\hbox{\it Vect}\,(M)}

\def\longbar#1{\setbox1=\hbox{$#1$}
\setbox2=\vbox{\hrule width 0.8\wd1}
\raise0.5\ht1\hbox{${\lower\dp1\box2}\atop\box1$}}  

\def\mediumbar#1{\setbox1=\hbox{$#1$}
\setbox2=\vbox{\hrule width 0.6\wd1}
\raise0.5\ht1\hbox{${\lower\dp1\box2}\atop\box1$}}

\begin{document}
\title{Dimer geometry, amoebae and a vortex dimer model} 
\author{Charles Nash}
\email[]{cnash@thphys.nuim.ie}
\affiliation{Department of Theoretical Physics, 
NUIM, Maynooth, Kildare, Ireland.}
\author{Denjoe O'Connor}
\email[]{denjoe@stp.dias.ie}
\affiliation{School of Theoretical Physics,
DIAS, 10 Burlington Road, Dublin 4, Ireland. }

\begin{abstract} We present a geometrical approach and introduce a
connection for dimer problems on bipartite and non-bipartite graphs.
In the bipartite case the connection is flat but has non-trivial ${\bf
  Z}_2$ holonomy round certain curves. This holonomy has the
universality property that it does not change as the number of
vertices in the fundamental domain of the graph is increased. It is
argued that the K-theory of the torus, with or without punctures, is
the appropriate underlying invariant.  In the non-bipartite case the
connection has non-zero curvature as well as non-zero Chern
number. The curvature does not require the introduction of a magnetic
field.  The phase diagram of these models is captured by what is known
as an amoeba. We introduce a dimer model with negative edge weights
which correspond to vortices.  The amoebae for various models are
studied with particular emphasis on the case of negative edge weights.
Vortices give rise to new kinds of amoebae with certain singular
structures which we investigate. On the amoeba of the vortex full
hexagonal lattice we find the partition function corresponds to that
of a massless Dirac doublet.
\end{abstract}

\maketitle
\section{Introduction}
\label{SecIntroduction}
The subject of dimers has a large literature and has attracted the
interest of both mathematicians and physicists. A few useful
mathematical and physical sources are
\cite{OkounkovKenyonSheffield,OkounkovKenyon2006,CimasoniReshetikhin_I,CimasoniReshetikhin_II,Broomhead:2012} and
\cite{Kasteleyn:1963,Fisher:1966,Nagle1989,Hanany-Kennaway:2005,Franco-et-al:2006,Hanany-Kennaway:2005, Feng-et-al:2008,OkounkovReshetikhinVafa:2006,Dijgraaf:2009}
respectively, as well as references therein.

Dimer partition functions can be expressed as a sum of Pfaffians of a
Kasteleyn matrix, $K$, which is a signed weighted adjacency matrix.
The dimer partition function, with uniform weights, counts the number of
perfect matchings of a graph. The model is naturally considered with
positive weights and can have a non-trivial phase diagram as the
weights are altered. In particular, it has a gapless phase which is
described by the amoeba of a certain curve known as the spectral curve
\cite{OkounkovKenyonSheffield,OkounkovKenyon2006} (see section
\ref{SecAmoebae}). The Kasteleyn matrix can be thought of as a discrete
lattice Dirac operator
\cite{CimasoniReshetikhin_I,CimasoniReshetikhin_II,Nash_OConnor_jphysa:2009}
and the finite size corrections to the partition function in the
scaling limit coincide with that of a continuum Dirac--Fermion on a
torus \cite{Nash_OConnor_jphysa:2009}. If one further adds signs to
the weights the model describes a lattice Dirac operator in a fixed
${\bf Z}_2$ gauge field background. The presence of additional signs
we refer to as the presence of vortices in the dimer system.

We review the basic construction of dimer models and give a
detailed construction of the connection on the determinant line bundle
over the positive frequency eigenvector space of the Kasteleyn matrix.
We find that 
\begin{itemize}
\item In the bipartite case the determinant line bundle has a flat
  connection.
\item The flat connection has non-trivial ${\bf Z_2}$ holonomy in
  accordance with the $\widetilde KO$-theory of the torus.
\item When vortices are included the system can describe additional
  massless Fermions and we present an example where the partition
  function, in the vortex full case of a hexagonal lattice,
  corresponds to a massless Dirac Fermion doublet.
\item For certain vortex configurations, the domain where the
  system describes a massless Dirac operator, the amoeba can develop a pinch.
\item The presence of vortices alters the thermodynamic phases, we
  exhibit a case where the gapped phase---an island, or compact oval in
  the amoeba corresponding to a massive Dirac phase---can, on introduction
  of vortices,  shrink and even disappear.
\end{itemize}

The paper is organised as follows: section \ref{SecDimers} describes
basic results on dimers and dimer partition functions, focusing on
bipartite dimer models. Sections \ref{SecAmoebae} and
\ref{SecPhasesAndAmoebae} describe a mathematical object known as an
amoebae which describes the gapless parameter domain of the Kasteleyn
matrix $K$.  In section \ref{SecDimerConnectionsAndCurvatures} we
construct the vector bundle of positive eigenvalues of $i K$,
and show that its determinant line bundle has a flat connection and in
section \ref{SecHolonomyAndFlatConnections} we show that this
connection has non-trivial ${\bf Z_2}$-holonomy in accordance with the
$\widetilde K O$-theory of the punctured torus. Section
\ref{SecNonbibartiteGraphsAndChernNumbers} treats dimers on
non-bipartite graphs.  Section \ref{SecNonHarnackAmoebae} is devoted to dimer models in 
to the presence of vortices. In section \ref{Concludingsec}  we present
our  conclusions;  this is followed by an appendix on some of the relevant K-theory.

\section{Dimers}
\label{SecDimers}
Dimer models are concerned with the set of vertex matchings of a graph,
or lattice, $\Gamma$. We shall consider $\Gamma$ to be a bipartite
or non-bipartite lattice with $2 n$ sites, or vertices, on the two
torus $T^2$. A perfect matching, $m$, on $\Gamma$ is a disjoint collection
of edges that contains  all the vertices: for $m$  to exist $\Gamma$ must have 
an even number of vertices. An edge belonging to a matching is called a 
{\it dimer} and perfect matchings are the same thing as dimer configurations.
 
We denote the set  of dimer configurations on $\Gamma$ by ${\cal M}(\Gamma)$.
Then to each matching, $m\in{\cal M}(\Gamma)$, we assign a
weight ${\rm e}^{-{\cal E}(m)}$; ${\cal E}(m)$ is normally required 
to be real in which case all weights are positive. We will find it useful
to go beyond this restriction in the latter part of this paper and
consider signed weights, but for the moment we take all weights to be positive.

Given this data
the {\it dimer partition function} 
$Z(\Gamma)$ is given by:
\begin{equation}
Z(\Gamma)=\sum_{m\in {\cal M}(\Gamma)}{\rm e}^{-{\cal E}(m)}\, .
\end{equation}
Each matching, $m$, consists of $n$ dimers with positive {\it edge weights},
$a_{e_1},\ldots,a_{e_n}$ whose relation to ${\cal E}(m)$ is that
where $e_i$ are the edges of the matching $m$, and $a_{e_i}$ is the weight
associated with the edge $e_i$, then 
\begin{equation}
{\rm e}^{-{\cal E}(m)}=\prod_{e_i\in m}a_{e_i}
\end{equation}
yielding 
\begin{equation}
Z(\Gamma)=\sum_{m\in {\cal M}(\Gamma)}\prod_{e_i\in m}a_{e_i}\, .
\end{equation}

More generally one can consider $p$ dimers: i.e. $p$ matched edges
where $p$ is less than or equal to the total number of edges; when
$p=n$ one has a dimer model, otherwise one has a monomer-dimer
system. Monomer-dimer systems have not yet yielded to exact solution
methods.

An alternative generalisation is to consider some of the weights being
negative, we will refer to such a system as containing vortices.  It
is relatively straightforward to solve for the partition function of such
systems and as we shall see they have a rich physics. We shall only
concern ourselves with dimer models, and dimer models with vortices,  in
this paper.

One can also investigate the probability of one, two, or more dimers (or 
edges), belonging to a matching $m$: to do this one uses,
the characteristic function  
$\sigma_{e_i}$ of an edge $e_i$,  and the  dimer-dimer
correlation functions $\left<\sigma_{e_1}\cdots \sigma_{e_k} \right>$: 
their joint definitions are   that 
\begin{equation}
\begin{aligned}
\sigma_{e_i}(m)&=\begin{cases}
               1 &\text{if } e_i\in m\\
               0 & \text{otherwise}\\
              \end{cases}\\
\left<\sigma_{e_1}\cdots \sigma_{e_k} \right>&= \frac{Z(e_1,\ldots, e_k;\Gamma)}{Z(\Gamma)}\\
\hbox{ with }\quad Z(e_1,\ldots, e_k;\Gamma)&=\sum_{\left\{m \mid e_1,\ldots e_k\in m\right\}} {\rm e}^{-{\cal E}(m)}\, . 
\end{aligned}
\end{equation}

For a planar region in  ${\bf R}^2$, and a graph $\Gamma$ with $2n$ vertices, Kasteleyn
  \cite{Kasteleyn:1963} showed that  
\begin{equation}
Z(\Gamma)=\vert \pfaff(K)\vert 
\end{equation}
where $\pfaff(K)$ is the Pfaffian of the 
Kasteleyn matrix $K$ which is a $2 n\times 2n$ 
{\it signed}, antisymmetric,  weighted adjacency matrix for $\Gamma$: Kasteleyn's sign assignments in $K$  are precisely what is needed to convert  $\pfaff(K)$ 
into the sum  of positive terms that constitute the partition function $Z(\Gamma)$. 

We specialise, for the moment, to the case where  $\Gamma$  is bipartite, 
so that we can  colour $n$ vertices  black and the other $n$ white. This allows 
us to write $K$ in the form 
\begin{equation}
K=\left(
\begin{matrix}
0 & \hat A\\
-\hat A^T & 0\\
\end{matrix}
 \right)
\end{equation}        
with $\hat A$  an $n\times n$ real matrix (we write $\hat A$ rather than $A$ since, later on, we want to use $A$ to denote a connection) and $T$ denotes transpose; note, too,  that $\pfaff K=(-1)^{n(n-1)/2}\det \hat A$.

 Let   $V^B={\bf C}^n$ and $V^W={\bf C}^n$ be vector spaces generated by the sets of black and white vertices  respectively, 
then  $\hat A$ is the linear  map   
\begin{equation}
\hat A:V^B\longrightarrow V^W
\end{equation}        
defined by  
\begin{equation}
\hat A_{ij}=\begin{cases}
\epsilon_{ij} a_{ij} &\text{if $i$ is connected to $j$ by
  an edge with weight $a_{ij}$}\\
               0 &\text{otherwise}\\
\end{cases}
\end{equation}        
where $\epsilon_{ij}=\pm1$ is the sign associated to the edge
whose weight is $a_{ij}$.

The signs $\epsilon_{ij}$ are computed by the 
 {\it clockwise odd rule} \cite{Fisher:1966}: arrows are placed on the edges  and $\epsilon_{ij}=+1$ when following an arrow 
and $-1$
when opposing one, and the product of the signs associated with any
fundamental plaquette is $-1$ when the plaquette is circulated in an
anticlockwise direction;  such an assignment of signs is called
a {\it Kasteleyn orientation}. 

Note that  all closed paths on bipartite graphs  have an even number
of edges: thus clockwise odd is also anticlockwise odd. 
This is false for non-bipartite graphs which, when Kasteleyn oriented,
possess an orientation which can be detected---cf. below  
where we discuss Chern numbers.   

The general result \cite{CimasoniReshetikhin_I,CimasoniReshetikhin_II} for
a graph $\Gamma$ embedded in a closed oriented surface $\Sigma$ of
genus $g$, is that the partition function of the dimer model on $\Gamma$
is given by
$$Z(\Gamma)=\frac{1}{2^g}\sum_{\vec{u}\in {\cal S}(\Sigma)}{\rm Arf}(\vec{u})\pfaff(K_{\vec{u}}(\Gamma))$$
where ${\cal S}(\Sigma)$ denotes the set of equivalence classes of
the $2^{2g}$ spin structures on $\Sigma$, ${\rm Arf}(\vec{u})=\pm 1$ is the
{\rm Arf}-invariant of the spin structure labeled by $\vec{u}$,
and $K_{\vec{u}}(\Gamma)$ is the Kasteleyn matrix with these boundary conditions.

When the graph $\Gamma$  is on a torus $T^2$,
the weighted sum is over the four different Kasteleyn matrices $K_{u,v}$ corresponding
to the four choices of periodic and anti-periodic boundary conditions around the cycles of the torus and 
\begin{equation}
K_{u,v}=\left(
\begin{matrix}
0 & \hat A_{u,v}\\
-\hat A^T_{u,v} & 0\\
\end{matrix}
 \right)\, .
\end{equation}
Each term in the sum corresponds to one of the $4$  discrete spin structures
of $\Gamma$ on $T^2$ and, writing  $Z({\rm e}^{2\pi i u},\rm{e}^{2\pi i v})=\det \hat A_{u,v}$,  one has 
\begin{equation}
Z(\Gamma)=\frac{1}{2}\left\{Z(1,-1)+Z(-1,1)+Z(-1,-1)-Z(1,1)\right\}\, .
\end{equation}

Note that since $\Gamma$ is on a torus it is doubly periodic and
can be realised as a quotient: one has 
\begin{equation}
\Gamma=\frac{\widetilde \Gamma}{{\bf Z}^2}
\end{equation}
where  $\widetilde \Gamma$ is the graph in the plane 
consisting of all possible translations by elements of ${\bf Z}^2$ of an
appropriate fundamental domain contained in $\Gamma$.  However, $\Gamma$ itself
may be  multiple copies of this fundamental domain where the weights
are repeated as translates. Then, by Fourier transforming $K_{u,v}$ we obtain
a matrix $K_{u,v}(z,w)$ with off-diagonal block $\hat A_{u,v}(z,w)$.

On the torus corresponding to the $N\times M$ translates of a 
fundamental tile, the spin structure decomposition for the partition function is given by\footnote{The overall sign
  ${(-1)}^{NM(NM-1)/2}$ ensures that the partition function is positive and is induced by the relation of $\pfaff K$ to $\det \hat A$. Both the sign and the
  leading expression here, in terms of Pfaffians,
  are valid for both bipartite and non-bipartite graphs. }
\begin{eqnarray}
Z=&&
{(-1)}^{NM(NM-1)/2}\frac{1}
     {2}\sum_{u,v=0}^{1/2}{\rm e}^{2\pi i(\frac{1}{2}+u+v+2uv)} 
\pfaff K_{u,v}\nonumber\\
&&=\frac{1}{2}\sum_{u,v=0}^{1/2}{\rm e}^{2\pi i(\frac{1}{2}+u+v+2uv)}\det \hat A_{u,v}
\end{eqnarray}

The determinant of $\hat A_{u,v}$ results in
a polynomial $\hat P(z,w)$ with 
\begin{eqnarray}
\det \hat A_{u,v}&&=\prod_{n=0}^{N-1}\prod_{m=0}^{M-1}
\hat P({\rm e}^{-\frac{2\pi i(n+u)}{N}},{\rm e}^{\frac{2\pi i(m+v)}{M}})\nonumber\\
\label{Det_uv}\ .
\end{eqnarray}
A key property of  the polynomial $\hat P$ is that it has a  pair of
zeros $(z_0,w_0)$, $(\bar z_0,\bar w_0)$ related by complex conjugation.

We now describe the construction of $\hat P(z,w)$. 
Let us label an arbitrary translated copy of the   
fundamental domain from which $\Gamma$ is built
by $\Gamma^{I,J}$ with $I,J\in{\bf Z}$.
Here  $I>0$ denotes the number of horizontal translations to 
the right, and $I<0$ the number of translations to the left,
the integer  $J$ labels vertical translations in a 
similar way; the fundamental domain is $\Gamma^{0,0}$. 

Thus one has 
\begin{equation}
  \Gamma=\bigcup_{I= 0}^{ N-1} \bigcup_{J= 0}^{M-1} \Gamma^{I,J}
  \quad\hbox{while}\quad \widetilde \Gamma=\bigcup_{I,J\in{\bf Z}}  \Gamma^{I,J} \, .
\end{equation}
Now to each $\Gamma^{I,J}$ we associate the matrix  $\hat A^{I,J}(z,w)$; so that, if $\epsilon_{ij} a_{ij}$  denotes the signed weight of an edge joining 
vertex $i$ to $j$, then 
we define 
\begin{equation}
\hat A^{I,J}_{ij}(z,w)=\begin{cases}       
               z^I w^J\sum_{edges}\epsilon_{ij} a_{ij}\quad + &z^L w^M\sum_{edges}\epsilon_{ij} a_{ij}              \\
               \text{(if edge lies in $\Gamma^{I,J}$) }
               &\text{(if edge crosses into an {\it adjacent} domain $\Gamma^{L,M}$)}\\
               0 &\text{otherwise.}\\
\end{cases}
\end{equation}     
With this data the fundamental domain containing the basic graph is 
$\Gamma^{0,0}$ and we define $\hat A(z,w)$ by writing  
\begin{equation}
\hat A(z,w)=\hat A^{0,0}(z,w)
\end{equation}
so that $\hat P(z,w)=\det \hat A^{0,0}(z,w)$ then  $\hat P(z,w)$ is a
Laurent polynomial in $z$ and $w$ with {\it real} coefficients. 

Note that if we defined $\hat A(z,w)$ using a domain other than the fundamental domain, 
e.g.\ if we wrote  $\hat A(z,w)=\hat A^{I,J}(z,w)$ then $\hat P(z,w)$ would just be multiplied by a monomial in $z$ and $w$ and the formula for the partition function would 
still hold true.  

Now both the infinite graph $\widetilde \Gamma$, and its bipartite black-white 
assignment,  are  translation invariant, and the decomposition 
\begin{equation}
\widetilde \Gamma=\bigcup_{I,J\in{\bf Z}}  \Gamma^{I,J} 
\end{equation}
can be viewed as  an indexing of $\widetilde \Gamma$  by the characters $z^I w^J$ of  the 
translation group ${\bf Z}^2$; in other words  $K^{I,J}(z,w)$ is simply the 
Fourier transform of the Kasteleyn matrix $\widetilde K$
of  $\widetilde \Gamma$. 

\begin{figure}
\begin{center}
\includegraphics[height=50mm]{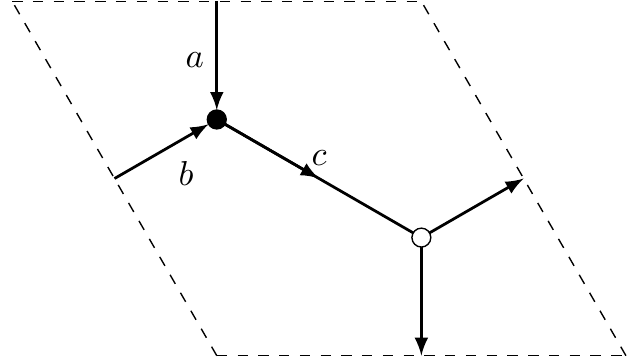}\qquad\qquad\qquad\qquad
\includegraphics[height=50mm]{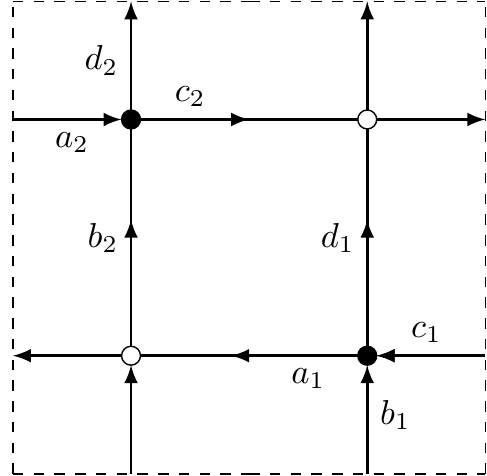}
\caption{A hexagonal and a rectangular example}
\end{center}
\label{twoexamples}
\end{figure}

A pair of  examples illustrating the above process can be quite simply given:
consider $\Gamma$ to be the graphs tiled by the fundamental domains
of figure \ref{twoexamples}.
Then, for the hexagonal graph,  we readily calculate that 
\begin{equation}
\begin{aligned}
             K(z,w)&=\left(
                       \begin{matrix}
                                   0 & c-b/z-a w\\
                        -(c-b z-a/w) & 0\\  
                        \end{matrix}
              \right)               \\
\Rightarrow \hat P(z,w) &=c-\frac{b}{z}-a \,w
\label{hexpolynomial}
\end{aligned}
\end{equation} 
while for the rectangular one we have 
\begin{equation}
\begin{aligned}
             K(z,w)&=\left(
                       \begin{matrix}
                                   0 & \hat A(z,w)\\
                        -\hat A^T(\bar z,\bar w) & 0\\  
                        \end{matrix}
              \right),    \qquad \hat A(z,w)=\left(
                                         \begin{matrix}
                                          a_1-c_1 z & d_1- b_1/w\\
                                          -b_2+ d_2 w & c_2 -a_2/z\\
                                         \end{matrix} 
                                         \right)         
\\
\Rightarrow \hat P(z,w) &=a_1 c_2+a_2 c_1 + b_1 d_2 + b_2 d_1 - \frac{a_1 a_2}{z} - c_1 c_2 \, z - d_1 d_2 \,w - \frac{b_1 b_2}{w}  \, .
\end{aligned}
\end{equation} 

Observe that varying the dimer weights $a$ and $b$ in (\ref{hexpolynomial})
is equivalent to moving $z$ and $w$ off the unit torus. In general,
for any polynomial arising from a bipartite dimer construction,  one
can always absorb combinations of dimer weights into $z$ and $w$ to move
them off the unit torus. 

\section{Amoebae}
\label{SecAmoebae}
Let  $F(z,w)=\sum_{i,j}c_{ij}  z^i w^j$  with $c_{ij},z,w\in{\bf C}$ be a polynomial, 
then its zero locus 
\begin{equation}
\begin{aligned}
F(z,w)&=0\\
(z,w)&\in ({\bf C^*})^2\\
\end{aligned}
\end{equation}
is a curve ${\cal C}$ in ${\bf C}^2$.    
The image of ${\cal C}$  in ${\bf R}^2$ under the logarithmic map ${\rm Ln}$ defined by
\begin{equation}
\begin{aligned}
&{\rm Ln}:{\cal C}\longrightarrow   {\bf R}^2\\
&\qquad (z,w)\longmapsto (\ln\vert z\vert,\ln\vert w\vert)\\
\end{aligned}
\end{equation}
is known as the  {\it amoeba}  ${\cal A(C)}$ of ${\cal C}$  so that 
${\cal A(C)}={\rm Ln}({\cal C})$. 

The polynomial  $\hat P(z,w)$ above is a special case of such
an $F(z,w)$ having {\it real coefficients} $c_{ij}$ and $\hat P(z,w)$  
is central to much of what follows; its zero locus ${\cal C}$ 
is known as the {\it spectral curve} of the graph $\Gamma$.  

\begin{figure}
\subfigure[]{\includegraphics[width=65mm]{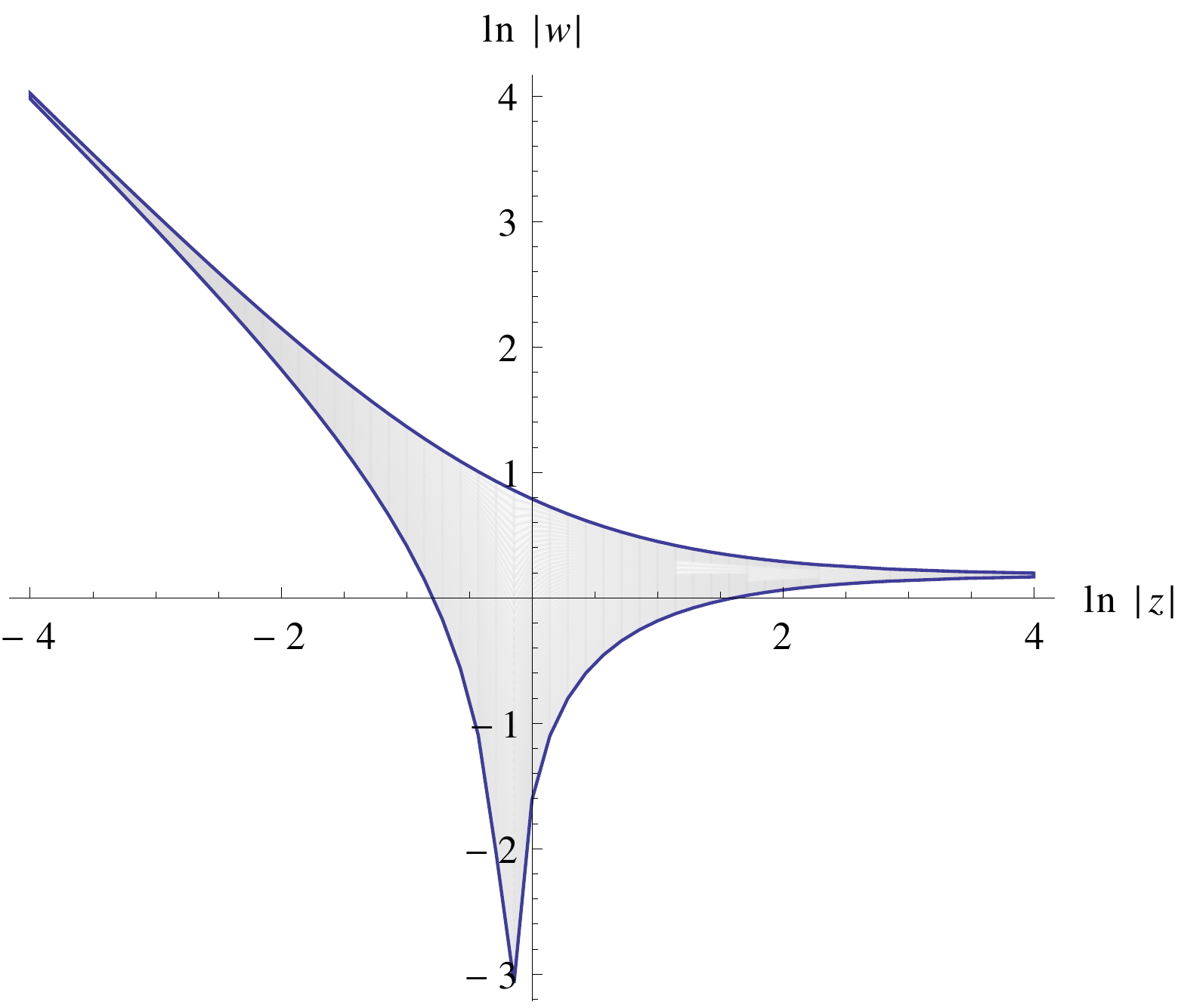}}
\hfill
\subfigure[]{\includegraphics[scale=0.4]{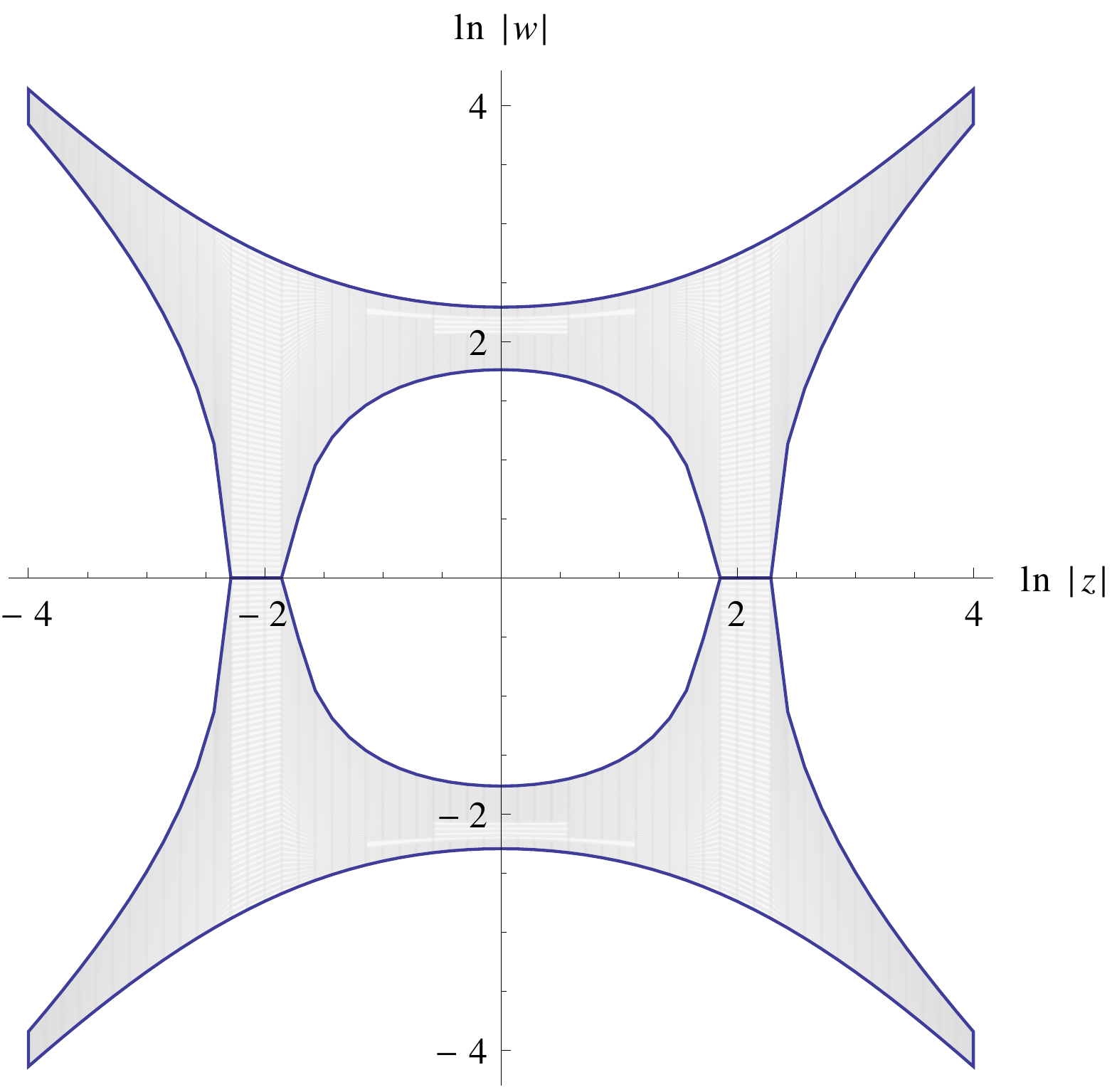}}
\caption{(a) Amoeba for $\hat P(z,w)=1.2-z^{-1}-w$ \hfill(b) Amoeba for $\hat P(z,w)=2+2\cosh(t)-z-z^{-1}-w-w^{-1}$,\, $\cosh(t)=2$}
\label{twoamoebae}
\end{figure}

The amoebae for the two graphs of figure  \ref{twoexamples} are
displayed in figure \ref{twoamoebae}; the closed curve in
figure \ref{twoamoebae} (b) is called a {\it compact oval}---we have moved off
the unit torus and used weights
$a=\vert w\vert$, $b=1/\vert z \vert$, $c=1.2$ for figure \ref{twoamoebae}  (a);
and for example $a_1={\rm e}^t$, $a_2^{-1}=\vert z\vert {\rm e}^t$, $d_1=1/b_2=\vert w\vert $
with $\cosh(t)=2$ and all remaining weights
unity for figure \ref{twoamoebae} (b).

Although an amoeba ${\cal A(C)}$  is unbounded in ${\bf R}^2$ it has 
finite area; in fact its area is bounded above by an irrational multiple of the area of the Newton polygon of $F(z,w)$---the Newton polygon $\Delta$ 
being the convex hull of the (integer) points $(i,j)\in{\bf R^2}$ for which $c_{ij}\not=0$. More precisely one has
\begin{equation}
{\rm Area\,}({\cal A(C)})\le \pi^2{\rm Area\,} (\Delta)\, .
\end{equation}

After multiplication of $\hat P(z,w)$ by a suitable 
monomial to eliminate its negative powers, one obtains a polynomial $\widetilde P(z,w)$ 
of degree $d$, say.  Since $\widetilde P(z,w)$  has real coefficients, it  
determines the real homogeneous polynomial $t^d \widetilde P(z/t,w/t)$  in $(z,w,t)$---where now $(z,w,t)\in{\bf R}^3$---and 
thus a real algebraic curve     
in ${\bf RP^2}$, this being  natural geometrical data possessed by  the spectral curve  ${\cal C}$.  We denote this real algebraic curve by $R{\cal C}$.  When plotting the  amoeba ${\cal A(C)}$  the signs of $z$ and $w$ play an essential role in
determining all its components which,  in turn, constitute the amoeba boundary. 

For dimer models on bipartite graphs $R{\cal C}$  is a {\it Harnack curve}.
Harnack curves are very  special curves possessing the maximal number of
components: i.e.  \mbox{$(d-1)(d-2)/2+1$}. The integer
\mbox{$(d-1)(d-2)/2$} is the genus $g$ of the curve and is equal
to the number of compact ovals of the amoeba. 

It will be  convenient for us to abuse terminology slightly and
often refer to a curve ${\cal C}$ (rather than $R{\cal C}$)  as being Harnack,
the context should prevent any confusion.

A fundamental result of Kenyon, Okounkov,
and Sheffield \cite{OkounkovKenyonSheffield} and Kenyon and Okounkov  
\cite{OkounkovKenyon2006} is that every Harnack curve arises in this
way so that the correspondence between the spectral curves of periodic
hexagonal dimer models and Harnack curves is a bijection. 
Also all bipartite planar graphs can be realised inside a
large enough hexagonal lattice by a combination of setting some dimer edge weights to zero
and bond contraction \cite{Fisher:1966,OkounkovKenyon2006};  
thus no bipartite graph is excluded. In addition, the most general
$d\times d$ hexagonal dimer model yields a generic
Harnack curve of genus $(d-1)(d-2)/2$.

Positive rescaling of $z$ and $w$ gives a free action
of ${\bf R^+}\times {\bf R^+} $ on the set ${\cal H}$ of Harnack curves;
if one quotients ${\cal H}$ by this action one obtains what is called
in \cite{OkounkovKenyon2006} the {\it moduli space of Harnack curves}.
Amoebae  provide natural coordinates for this moduli space: these
coordinates being the  areas of the holes and the distances between
the tentacles \cite{OkounkovKenyon2006}.

When a  curve ${\cal C}$ is Harnack  the area of its  amoeba  is
maximal and saturates  the area inequality above---i.e.  
\begin{equation}
{\rm Area\,}({\cal A(C)})=\pi^2{\rm Area\,} (\Delta)
\end{equation}
for ${\cal C}$ Harnack. 

The converse of this equality also holds in the sense
that ${\rm Area\,}({\cal A(C)})=\pi^2{\rm Area\,}(\Delta)$ 
implies that the curve ${\cal C}$ of $F(z,w)$
(after possible rescaling of $z$, $w$ and $F$ by complex constants) is
invariant under complex conjugation and possesses a real part
which determines a Harnack curve
$R{\cal C}$  in ${\bf RP^2}$---cf. \cite{Mikhalkin_Rullgard2001}
for more details. 

Since the polynomial $\hat P(z,w)$ has real coefficients,
the  map ${\rm Ln}:{\cal C}\longrightarrow   {\bf R}^2$ is generically,
at least $2$ to $1$; however when ${\cal C}$ is Harnack ${\rm Ln}$
is  $2$ to $1$ everywhere, except at real nodes which occur on the
boundary of ${\cal A(C)}$,
cf. \cite{OkounkovKenyon2006},\cite{Mikhalkin_Rullgard2001}.
In addition $\hat P(z,w)$ has exactly two zeroes,
$p=(z_0,w_0)$ and $\bar p=(\bar z_0,\bar w_0)$, on the unit torus $T^2$. 

\section{Phases and amoebae}
\label{SecPhasesAndAmoebae}
The amoeba can be viewed as the massless or gapless phase with its bounding
curves as the phase boundaries in a dimer model phase
diagram \cite{OkounkovKenyonSheffield}: the complement of the amoeba
consists of both compact and non-compact regions.  In the terminology of
dimer models, as models of melting crystals, the non-compact regions
exterior to the bounding ovals constitute the frozen regions. The amoeba
itself is referred to as the liquid phase, and the interior of the 
compact ovals as the gaseous phase. There are also useful applications
of these ideas to the Kitaev model \cite{Nash:2009prl} and
topological phase transitions \cite{Nash_OConnor_jphysa:2009}.

These different phases arise naturally when one calculates the
correlation functions between the edges of $\Gamma$. The correlation functions 
possess three types of decay \cite{OkounkovKenyonSheffield}---where a
decay is measured by the fall off of
$<\sigma_{e_1}\sigma_{e_2}>-<\sigma_{e_1}><\sigma_{e_2}>$ with
distance between $e_1$ and $e_2$---these types being exponential,
polynomial, or no decay, and they correspond to the gaseous, liquid and
frozen phases respectively. In the context of Kasteleyn matrices as
Dirac operators, the amoeba is the massless phase while the interiors
of the compact ovals correspond to massive Dirac operators
\cite{Nash_OConnor_jphysa:2009}.

\section{Dimer connections and curvatures}
\label{SecDimerConnectionsAndCurvatures}
Now let $(z,w)=({\rm e}^{i\theta},{\rm e}^{i\phi})$ once again denote coordinates on $T^2$, rather than on ${\bf C^2}$, and let   
 $\Gamma$  be a bipartite graph on  $T^2$  whose Kasteleyn matrix $K$, when Fourier transformed, gives $K(z,w)$  where  
\begin{equation}
K(z,w)=\left(
\begin{matrix}
0 & \hat A(z,w)\\
-\hat A^T(\bar z,\bar w) & 0\\
\end{matrix}
 \right)_{2n\times 2n} \, .
\end{equation}
With these conventions the matrix $i K(z,w)$ is Hermitian. Here we use
$2n$ for the number of vertices in the fundamental tile, in contrast
to the usage in the introduction where $2n$ referred to the total
number of vertices in the graph $\Gamma$.

If $\Gamma$ is {\it non-bipartite} its Kasteleyn matrix  $K$ also has a Fourier transformed $(0,0)$ component $K(z,w)$ of the form
\begin{equation}
K(z,w)=\left(
\begin{matrix}
D_1(z,w) & \hat A(z,w)\\
-\hat A^T(\bar z,\bar w) & D_2(z,w)\\
\end{matrix}
 \right)_{2n\times 2n}
\end{equation}
with at least one of the diagonal blocks $D_1(z,w)$ and $D_2(z,w)$ being non-zero and such that $i K(z,w)$ is still Hermitian.

We now describe how to use this data to  construct a certain connection  on 
$T^2$. The $n$-dimensional space of eigenvectors of  $i K(z,w)$ with positive 
eigenvalues  form a rank $n$ bundle over $T^2$  which we denote by $E^+$.

Let $x\in T^2$, then the fibre $E^+_x$, at $x$,  has a basis 
consisting of the corresponding $n$ positive eigenvectors which we denote by  
\begin{equation}
v_1(x), v_2(x),\ldots v_n(x) \, .
\end{equation}
With respect to the standard complex inner product, fixed as  $x$ varies, let each 
eigenvector $v_i(x)$ have unit norm and,  
in an orthonormal basis, have components $v_{ji},j=1,\ldots N$. 
When taken together the  $v_i(x)$  form the non-square matrix $v(x)$ where 
\begin{equation}
v(x)=\left(\begin{matrix}
                     v_{11}(x)&\cdots & v_{1n}(x)\\
                    \vdots &\cdots & \vdots\\ 
                     v_{N1}(x)&\cdots &v_{N n}(x)\\ 
          \end{matrix}   
    \right)_{N\times n}\qquad(N=2n)
\end{equation}
giving one a map 
\begin{equation}
v(x):{\bf C}^n\longrightarrow {\bf C}^N \, .
\end{equation}
As $x$ varies the map $v(x)$  embeds the fibres $E^+_x$ of $E^+$---and thus the 
whole bundle---in the trivial bundle $T^2\times {\bf C}^N$; 
conversely, if  $v^*(x)$ is the adjoint of $v(x)$,  
the map $P=v(x) v^*(x)$ is an orthogonal projection from ${\bf C}^N$ to ${\bf C}^N$ on which rests the non-triviality of $E^+$. 

Summarising, and abbreviating $v(x)$ and $v^*(x)$ by $v$ and $v^*$ 
respectively, yields  
\begin{equation}
\begin{aligned}
v&=\left(\begin{matrix}
                     v_{11}(x)&\cdots & v_{1n}(x)\\
                    \vdots &\cdots & \vdots\\ 
                     v_{N1}(x)&\cdots &v_{N n}(x)\\ 
          \end{matrix}   
    \right)_{N\times n}\\
 &\\
P&=v v^* \\
\end{aligned}
\quad\begin{aligned}
v^* &=\left(
                 \begin{matrix} 
                           \bar v_{11}(x)&\cdots &\bar v_{1N}(x)\\
                           \vdots &\cdots & \vdots\\
                    \bar v _{n1}(x)&\cdots & \bar v_{n N}\\
                 \end{matrix}  
       \right)_{n\times N} \\
         &\\ 
v^* v&=I_{n\times n} \\
\end{aligned}
\qquad (N=2n)
\end{equation}
where $I_{n\times n}$ denotes the identity matrix on ${\bf C}^n$ and  $P^2=P$; 
$v$ is called a {\it partial isometry}---it is not a 
real isometry   since $v$ is not a square matrix. 

A section $s$ of $E^+$ is then a map taking values
in ${\bf C}^N$---i.e. one has 
\begin{equation}
s:T^2\longrightarrow E^+_x \subset {\bf C}^N \, .
\end{equation}
However, as usual, derivatives of $s$ such as $\partial_\mu s$ may not,
as $x$ varies, still be $E^+_x$-valued, but we can project them 
back onto $E^+$ to take care of this problem thereby creating
a covariant derivative on $E^+$. Thus the covariant derivative of 
$s$ is $\nabla_\mu s$ where 
\begin{equation}
\nabla_\mu s=P\,\partial_\mu s \, .
\end{equation}

Our choice of inner product above means that $\nabla_\mu$ is the
covariant derivative corresponding to a $U(n)$
connection $A=A_\mu dx^\mu$, say, which we 
can identify by direct calculation as follows: if  $f$ is a map 
\begin{equation}
f:T^2\longrightarrow {\bf C}^n
\end{equation}
then the product $v f$ gives us the section 
\begin{equation}
T^2\xrightarrow{\;f\;}{\bf C}^n\xrightarrow{\;v\;}{\bf C}^N
\end{equation}
and the covariant derivative formula gives 
\begin{equation}
\begin{aligned}
\nabla(v f)&= P d(v f)\\
                      &=v v^*  d(v f) \\
                      &= v v^* \left\{d v f +v d f\right\}\\
                      &=v d f+ v v^*  d v f\\
&=v\left(d f +A f\right),\quad\hbox{where }\quad A=v^* d v \\
                      &=\nabla (vf),\qquad\quad\;\;\hbox{with }\quad\;\;\nabla=v(d+A)v^* 
\end{aligned}
\end{equation}
Hence the connection is the $U(n)$ matrix $A$ where 
\begin{equation}
A=v^* d v  
\end{equation}
and its curvature is
\begin{equation}
            d A+ A \wedge A =d v^* \wedge d v+v^* d v \wedge v^* d v 
\end{equation}
So the covariant derivative and curvature on $E^+$ are
$\nabla$ and $F$ respectively, with $F=\nabla^2=v(d A+ A \wedge A)v^*$. A routine calculation shows that
\begin{equation}
  F=P\, d P \wedge d P P
\end{equation}

We shall examine the connection and its curvature in the subsequent sections but note that its introduction has not required the presence of a magnetic field. 
\section{Holonomy and Flat connections}
\label{SecHolonomyAndFlatConnections}
We will be interested in the $U(1)$ connection associated with $tr (F)$ as it
is this that gives the first Chern class of the connection. 
It turns out that for {\it bipartite graphs} the curvature $tr(F)$ is zero
so that the associated $U(1)$ connection is flat: nevertheless
this connection is non-trivial as it has non-trivial holonomy as we shall now show.
\begin{figure}
\begin{center}
\includegraphics[height=50mm]{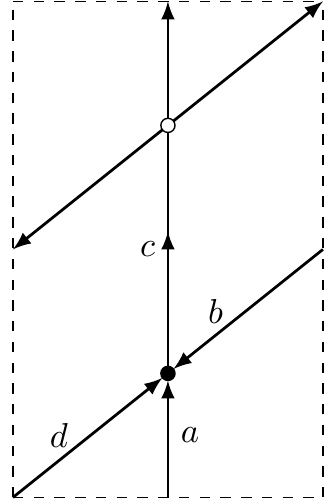}
\caption{A bipartite graph with $\hat P(z,w)=c-b\, z- \frac{a}{w}- \frac{d}{w z}$}
\end{center}
\label{twositebipartite}
\end{figure}

It is instructive to first study a case where $n=1$ and we do this for the graph $\Gamma$ shown in figure \ref{twositebipartite}.
For the bipartite case, when $D_1=D_2=0$, one  punctures the torus $T^2$ at the points $p$ and $\bar p$  since $\hat P(z,w)$ vanishes there, we denote the punctured torus by $T^2_{p,\bar p}$. One   
finds that the connection $A$ and its curvature $F$ are given by 
\begin{equation}
\begin{aligned}
A&=\frac{1}{2}f^{-1} df,\quad \text{ with }\quad f=\sqrt{\frac{\hat P(z,w)}{\overline{\hat P(z,w)}}}\\
F&=0\\
\end{aligned}
\end{equation}
so that we have a flat connection; however the connection $A$ is {\it not trivial} as it has 
non-trivial holonomy    $\exp\left[\int_C A\right]$ for some curves $C$ on $T^2_{p,\bar p}$. In other words for such curves

\begin{equation}
\begin{aligned}
\exp\left[\int_C A\right]&\not=1\\
\text{i.e.} \qquad \int_C \frac{A}{2\pi i}&\not\in{\bf Z}\\
\end{aligned}
\end{equation}    

As an example, for figure \ref{twositebipartite} choose $a=b=c=d=1$
so that  
\begin{equation}
\hat P(z,w)=1-\frac{1}{w}-z-\frac{1}{wz}
\end{equation}
and thus $\hat P(z,w)=0$ at the points 
\begin{equation}
p=({\rm e}^{i \pi/2},1),\qquad 
\bar p =({\rm e}^{-i \pi/2},1)
\end{equation}  
One then immediately discovers by direct calculation that if $C$ is a small circle  
\begin{equation}
\int_{C}\,\frac {A}{2\pi i}=\begin{cases} 
                                        \;\;\;\frac{1}{2} & \text{if  $C$ encircles $p$ but not $\bar p$}\\ 
                                             -\frac{1}{2} & \text{if  $C$ encircles $\bar p$ but not $p$}\\
                            \end{cases}
\end{equation}
 Hence we obtain non-trivial holonomy and it is easy to choose a different  $C$ and obtain  other  results: indeed if $C$ does not  contain $p$ or $\bar p$, 
but is non-contractible because it is a non-trivial homology cycle, then $\exp\left[\int_C A\right]$ can also have non-trivial holonomy.

For example if $C$ is the curve  $\theta=\theta_0$, $\theta_0$ constant, then  
\begin{equation}
\int_C \frac{A}{2\pi i}=\left\{
                                  \begin{matrix} -\frac{1}{2}&\text{if $\theta_0\in\left(-\frac{\pi}{2},\frac{\pi}{2}\right)$}\\
                                                             &\\
                                                  \quad 0          & \text{otherwise}\\
                                   \end{matrix}
                           \right.
                      \end{equation}
Non-trivial flat connections  require $T^2_{p,\bar p}$ to have a non vanishing fundamental group but one knows 
that $T^2_{p,\bar p}$ is homotopic to a bouquet of 
three circles (meaning three circles sharing one common point)   and therefore 
\begin{equation}
\pi_1(T^2_{p,\bar p})={\bf Z}*{\bf Z}*{\bf Z}\quad\left(\text{the free group on three generators}\right)
\end{equation}
so all is satisfactory. 

 These properties  of flatness and non-trivial holonomy persist---for the appropriate curvature and connection---when the bipartite graph  is enlarged as we now demonstrate.   

First let $F$ be the curvature coming from the Kasteleyn matrix $K(z,w)$  for a general  bipartite graph. 
 One has  
\begin{equation}
K(z,w)=\left(
\begin{matrix} 0&\hat A(z,w)\\
                       -\hat A(\bar z, \bar w)&0\\
\end{matrix}
\right)_{2 n\times 2n}
\end{equation}
with curvature $F=P dP\wedge dP P$ 
then it is easy to see that
\begin{equation}
\tr(F)=0\, .
\end{equation}
For, abbreviating  $\hat A(\bar z, \bar w)$ to $\hat A^*$, and setting $Q=(\hat A \hat A^*)^{-1/2}\hat A$,
$P$ can be written\footnote{For a (necessarily real)  eigenvalue $\lambda$ of $i K$ 
  with $i K v =\lambda$, and $v$ the column vector $v=(v_1,v_2)$,
  we find that
  $$
  \begin{matrix}
    i \hat A v_2=\lambda v_1, \hbox{ and }
    -i \hat A^* v_1=\lambda v_2 \\ 
   \Rightarrow  i \hat A \hat A^* v_1=\lambda^2 v_1\\  
   \Rightarrow    i \sqrt{\hat A \hat A^*} v_1=\vert \lambda \vert v_1\\ 
                                     \end{matrix}
  $$
  Now dividing our first displayed equation by $\sqrt{\hat A \hat A^*}$,  we deduce that, if $Q=({\hat A \hat A^*})^{-1/2} \hat A  $, then  
  $$
  i Q v_2=\frac{\lambda}{\vert\lambda\vert }v_1=\left\{\begin{matrix}
v_1&\hbox{ if } \lambda>0\\ 
-v_1&\hbox{ if } \lambda<0\\ 
\end{matrix}\right.\quad\hbox{ and }\quad  -i Q^* v_1=\frac{\lambda}{\vert\lambda\vert }v_2=\left\{\begin{matrix}
v_2&\hbox{ if } \lambda>0\\ 
-v_2&\hbox{ if } \lambda<0\\ 
\end{matrix}\right.
  $$
 so that  $Q Q^*= Q^* Q=I$, whence $P$ as given in \ref{projectionform} is the desired projection onto the positive eigenspace; $P$ also satisfies $[K,P]=0$. 
} as
\begin{equation}
\label{projectionform}
P=\frac{1}{2}\left(
            \begin{matrix}
                    I&iQ\\
                   -iQ^*&I\\
            \end{matrix}
            \right)_{2n\times 2n} 
\end{equation}
with  $I$  the $n\times n$ identity matrix.  Hence 
\begin{equation}
\begin{aligned}
dP&=\frac{1}{2}\left(\begin{matrix} 
                           0&i dQ\\
                            -i dQ^*&0\\ 
                   \end{matrix}  
             \right)\\
\Rightarrow dP\wedge dP&=\frac{1}{2^2}\left(\begin{matrix}
                                               dQ\wedge dQ^*&0\\
                                               0&dQ^*\wedge dQ\\
                                            \end{matrix}
                                                                        \right) \\
\end{aligned}
\end{equation}

and so we have,   
\begin{equation}
\begin{aligned}
\tr(F)&=\tr(P dP\wedge dP P)=\tr(P dP\wedge dP)\\
      &=\frac{1}{2^3}\tr\left\{\left(\begin{matrix}
                                        I&iQ\\
                                         -iQ^*&I\\
                                                             \end{matrix}
                                                      \right) 
                                                               \left(\begin{matrix}
                                                                      dQ\wedge dQ^*&0\\
                                                                      0&dQ^*\wedge dQ\\
                                                               \end{matrix}                                                                                                        \right)
                                                  \right\}\\
                              &=\frac{1}{2^3}\tr      \left(
                                                             \begin{matrix}
                                                                 dQ\wedge dQ^*& -i QdQ^*\wedge dQ\\
                                                               i Q^* dQ\wedge dQ^*& dQ^*\wedge dQ\\
                                                              \end{matrix}
                                                     \right)\\
                              &=\frac{1}{2^3}\left\{\tr\left(dQ\wedge dQ^*\right)+\tr\left(dQ^*\wedge dQ \right)\right\}\\
                              &=0\\
\end{aligned}
\end{equation}
as claimed.

Now, just as when $n=1$,  there is non-trivial holonomy  when $n>1$: we shall also find the interesting result that the 
holonomy obtained is universal and is independent of $n$.  

However first we must  identify  an appropriate    
line bundle $L\subset E^+$  with connection and to this end we shall use the following notation:  we   denote the connection and curvature  
on any bundle $E$ by  $F^E$ and $A^E$ respectively. 

So taking our bundle $E^+$---whose connection and curvature were formerly 
denoted by $A$ and $F$ above---we now denote these quantities by  
\begin{equation}
A^{E^+}
\end{equation}
and $F^{E^+}$ respectively. As we will see below, the  line bundle $L$ that we seek is simply the 
determinant line bundle of $E^+$, that is 
\begin{equation}
L=\det(E^+)
\end{equation}
Let $v_+^1,v_+^2\cdots v^n_+$ denote  the $n$ {\it unit normalised} positive eigenvectors of $i K(z,w)$, then this bundle has projection $P_{\det(E^+)}$ where 
\begin{equation}
\begin{aligned}
P_{\det(E^+)}&=V_{\det(E^+)} V^*_{\det(E^+)}\\ 
\text{ with}\qquad V_{\det(E^+)}&=v^1_+\wedge v^2_+\wedge \cdots \wedge v^n_+\\ 
\end{aligned}
\end{equation}
and the associated connection is therefore $A^{\det(E^+)}$ where
\begin{equation}
\begin{aligned} 
A^{\det(E^+)}&=V_{\det(E^+)}^*d V_{\det(E^+)}\\
             &=\left<v^1_+\wedge v^2_+\wedge \cdots \wedge v^n_+, d(v^1_+\wedge v^2_+\wedge \cdots \wedge v^n_+)\right>\\
\end{aligned}
\end{equation}
Here it may be useful to recall that, if $W$ is a vector space,  the inner product on $ \Lambda^n W$, which for orthonormal ${\bf e_i}\in W$ renders the vectors 
${\bf e_{i_1}}\wedge{\bf e_{i_2}}\wedge \cdots\wedge {\bf e_{i_n}}$  orthonormal, is given by 
\begin{equation}
\begin{aligned}
&<a_1\wedge a_2\cdots\wedge a_n ,b_1\wedge b_2\cdots\wedge b_n>=\det(M(a,b))\\ 
&M(a,b)=\left(
                 \begin{matrix}<a_1,b_1>&\cdots&<a_1,b_n>\\
                               <a_2,b_1>&\cdots&<a_2,b_n>\\
                                   \vdots &\ddots &\vdots\\
                              <a_n,b_1>&\cdots& <a_n,b_n>\\
                  \end{matrix}          
        \right)\\
\end{aligned}
\end{equation}
 In fact,
\begin{equation}
A^{\det(E^+)}=\tr(A^{E^+})
\end{equation}
To see this note first that 
\begin{equation}
\begin{aligned}
A^{\det(E^+)}&=<v_+^1,d v_+^1>+\cdots+<v_+^n,d v_+^n>\\ 
             &=(v_+^1)^* d v_+^1+\cdots+(v_+^n)^* d v_+^n\\
\end{aligned}
\end{equation}
and for $E^+$ we observe that $v_+^i$ is a $2n\times 1$ column vector and $(v_+^i)^*$ is a $1\times 2n$ row vector with  
\begin{equation}
\begin{aligned}
V_{E^+}&=\left(\begin{matrix}
                v_+^1&v_+^2&\cdots&v_+^n\\
                \end{matrix}
         \right)_{2n\times n}\\
V^*_{E^+}&=\left(\begin{matrix}
                  (v_+^1)^*\\(v_+^2)^*\\\vdots\\(v_+^n)^*\\
                  \end{matrix}
            \right)_{n\times 2n}\\
\end{aligned}
\end{equation}
where $V^*_{E^+} V_{E^+}=I_{n\times n}$ and  the projection $P_{E^+}$
is given by  $P_{E^+}=V_{E^+} V^*_{E^+}$. 
Hence 
\begin{equation}
A^{E^+}=V^*_{E^+} d V_{E^+}=\left(\begin{matrix}
                             (v_+^1)^*d v_+^1&\cdots&(v_+^1)^*d v_+^n\\
                                             \vdots&\cdots&\vdots\\
                                             (v_+^n)^*d v_+^1&\cdots&(v_+^n)^*d v_+^n\\ 
                                   \end{matrix}
                                    \right) 
\end{equation}
yielding
\begin{equation}
\tr(A^{E^+})=(v_+^1)^*d v_+^1+(v_+^2)^*d v_+^2+\cdots+(v_+^n)^*d v_+^n
\end{equation}
as claimed. 

One can check that $A^{\det(E^+)}=\tr(A^{E^+})$ is indeed a $U(1)$ connection; 
  it is also flat since its curvature $F^{\det(E^+)}$ 
satisfies
\begin{equation}
\begin{aligned}
F^{\det(E^+)}&=d\tr(A^{E^+})\\
             &= \tr(d_A A^{E^+})=\tr(F^{E^+})\\
             &=0\\
\end{aligned}
\end{equation}

Now we   are ready to 
calculate the holonomy  of $A^{\det(E^+)}$ round some curve $C$: let 
\begin{equation}
K=\left( \begin{matrix}
            0& \hat A\\
         -\hat A^* &0\\
         \end{matrix}
\right)
\end{equation}
and $u_+$ satisfy    
\begin{equation}
\hat A^* \hat A u_+=\lambda_+ u_+,\qquad \innprod{u_+}{u_+}=1
\end{equation}
then  $v_+$ is a unit norm  eigenvector of  $iK$ with eigenvalue $\sqrt{\lambda_+}$ where 
\begin{equation}
v_+=\frac{1}{\sqrt{2}}\left(
                            \begin{matrix}
                             \textstyle w_+\strut \\
                    {\textstyle  -i u_+}\\
                             \end{matrix} \right),\qquad w_+= \frac{\hat A u_+}{\sqrt{\lambda^i_+}}
\end{equation}
  
Using this orthogonal decomposition for each $v_+^i$ we calculate that 
\begin{equation}
\begin{aligned}
\left<v^1_+\wedge v^2_+\wedge \cdots \wedge v^n_+, d(v^1_+\wedge v^2_+\wedge \cdots \wedge v^n_+)\right>&=
\frac{1}{2}\left\{(u_+^1)^* d u_+^1+\cdots+(u_+^n)^* d u_+^n\right.\\
&\left.\quad+(w_+^1)^* d w_+^1+\cdots+(w_+^n)^* d w_+^n\right\}\\
=&\frac{1}{2}\left\{\left<w^1_+\wedge w^2_+\wedge \cdots \wedge w^n_+, d(w^1_+\wedge w^2_+\wedge \cdots \wedge w^n_+)\right>\right.\\
 &\;\;+\left.\left<u^1_+\wedge u^2_+\wedge \cdots \wedge u^n_+, d(u^1_+\wedge u^2_+\wedge \cdots \wedge u^n_+)\right>\right\}\\
\end{aligned}
\end{equation}

But now take note that 

\begin{equation}
\begin{aligned}
w^1_+\wedge w^2_+\wedge \cdots \wedge w^n_+&=\frac{\det(\hat A)}{
                    \sqrt{\lambda_+^1\lambda_+^2\cdots \lambda_+^n}}u^1_+\wedge u^2_+\wedge \cdots \wedge u^n_+\\
                                           &=f(z,w)\,u^1_+\wedge u^2_+\wedge \cdots \wedge u^n_+,\quad \hbox{ where }f(z,w)=
\sqrt{\frac{\hat P(z,w)}{\overline{\hat P(z,w)}}}\\
\end{aligned}
\end{equation}
and hence we obtain 
\begin{equation}
A^{\det(E^+)}=\frac{1}{2}f^{-1}d f+(u_+^1)^* d u_+^1+\cdots+(u_+^n)^* d u_+^n
\label{AdetE+}
\end{equation}
Thus our formula for the holonomy of $A^{\det(E^+)}$ round a curve $C$ on the torus $(z,w)=({\rm e}^{i\theta},{\rm e}^{i\phi})$ is  
\begin{equation}
\int_C A^{\det(E^+)}=\frac{1}{2}\int_C f^{-1}d f +\int_C (u_+^1)^* d u_+^1+\cdots+\int_C (u_+^n)^* d u_+^n
\end{equation}
Now let us take a basis for the $n$ dimensional space  on which $\hat A^* \hat A$ acts in which $u_+^i$ takes the form 
\begin{equation}
u_+^i=\left(\begin{matrix}0\\
                          \vdots\\ 
                           {\rm e}^{i\psi_i}\\ 
                           \vdots\\
                           0 \\
            \end{matrix}
       \right)_{n\times 1}\qquad                  \begin{matrix}
                                                          \psi_i\equiv \psi_i(\theta,\phi)\hfill\\
                                                          \psi_i(\theta,\phi)\hbox{ periodic in $\theta$ and $\phi$}\\
                                                   \end{matrix} 
\end{equation}
where the only non-zero entry in the representation of  $u_+^i$ above is $\psi_i$ which occupies  the $i^{th}$ position.   
This means that
\begin{equation}
\frac{1}{2\pi i}\int_C (u_+^i)^* d u_+^i\in{\bf Z}
\end{equation}
for each $i$. 
One can now easily check that
\begin{equation}
\frac{1}{2\pi i}\int_C A^{\det(E^+)}=                    \frac{1}{2}+k ,\,k\in{\bf Z}\text{$\quad$  if $C$ contains  one of the  zeroes of $\hat P(z,w)$ }
\end{equation}
The holonomies for other choices of $C$ can also be readily verified.

Thus the only non-trivial contribution to the holonomy group element comes from the term ${1/2}\int_C f^{-1}d f$
which is precisely the term obtained in the two site case:  we have therefore deduced  that
\begin{equation}
  \exp\left[\int_C A^{\det(E^+)}\right]=\exp\left[\frac{1}{2}\int_C f^{-1}d f\right]
  \label{holonomy_detE+}
\end{equation}
Hence, as claimed above, we have shown that, when we have $2n$ sites,
the holonomy of the connection $A^{\det(E^+)}$ is a universal
invariant independent of $n$. Notice that if the $\frac{1}{2}$ in
(\ref{holonomy_detE+}) is replaced by $1$ the holonomy becomes
trivial. This is precisely what occurs if we consider the matrix direct sum
of $K$ with itself, $K\oplus K$, whose bundle of positive eigenvectors
is $E^+\oplus E^+$ where, following (\ref{AdetE+}), we find that 
\begin{equation}
 \exp\left[ \int_C A^{\det(E^+\oplus E^+)}\right]=\exp\left[\int_C f^{-1}df\right]=1 
\end{equation}
and the holonomy is trivial.  This latter ${\bf Z_2}$ observation is a
$\widetilde KO$-theory one, see the appendix.

Passing to the underlying real bundles,  the real K-theory of $T^2_{p,\bar p}$---cf. the appendix---is given by
\begin{equation}
  \widetilde KO(T^2_{p,\bar p})={\bf Z_2}\oplus {\bf Z_2}\oplus {\bf Z_2}
\end{equation}  
and this captures the ${\bf Z_2}$-Fermionic nature of the holonomy for
bundles of any rank.  We further note that were we to perform the same
computation as above for the continuum massless Dirac operator on the plane we
would also find that the associated connection had ${\bf Z_2}$
holonomy. The relation with $\widetilde KO$ theory can be understood
in a similar manner, since, topologically, the punctured plane is $S^2_{p,q}$, 
the twice punctured sphere, and from the appendix we can deduce
that $\widetilde KO(S^2_{p,q })={\bf  Z_2}$.
One can conclude that it is the Fermionic nature of the dimer
system that is responsible for this topological observation.

\section{Non-bipartite graphs and Chern numbers}
\label{SecNonbibartiteGraphsAndChernNumbers}   
\begin{figure}
\begin{center}
\includegraphics[height=50mm]{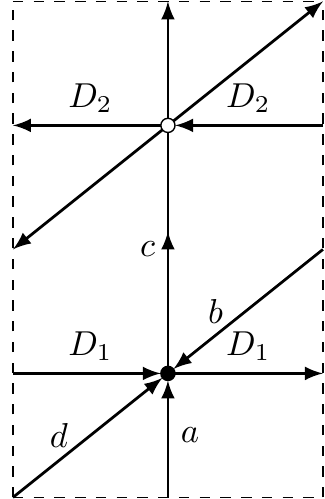}
\caption{A non-bipartite graph whose associated curvature is non-zero.}
\end{center}
\label{twositenonbipartite}
\end{figure}

Now we come to cases where the curvature is non-zero: this  happens
for non-bipartite graphs.
Such a case  occurs when, for example, two non-bipartite edges are
added to the graph of figure \ref{twositebipartite}. We display the
resulting graph $\Gamma$ in figure \ref{twositenonbipartite}. 

The  Kasteleyn matrix $K(z,w)$ of this $\Gamma$  is given by

\begin{equation}
\begin{aligned}
K(z,w)&=\left(\begin{matrix}
                  D_1(z-\bar z)&\hat P(z,w)\\
                 -\hat P(\bar z,\bar w)&-D_2(z-\bar z)\\
              \end{matrix}\right)\\
            \text{with }\qquad\qquad\quad
            \hat P(z,w)&=c-b\, z- \frac{a}{w}-\frac{d}{w z},\;D_1, D_2 \text{ constants}\\
\label{nonbipartitetwosite}
\end{aligned}      
\end{equation}

We   decompose $K(z,w)$ using the three Pauli matrices 
$\sigma_1,\sigma_2,\sigma_3$ and the identity, $I$, which,
for convenience, we denote by $\sigma_0$. This yields 
\begin{equation}
\begin{aligned}
i K(z,w)&=m_0\,\sigma_0 +m_1\,\sigma_1+m_2\,\sigma_2+m_3\,\sigma_3\\                                                                                        
        &=\left(\begin{matrix}
                m_0+m_3   & m_1-i m_2\\ 
                         m_1+i m_2 &m_0 -m_3 \\
                \end{matrix}\right),\quad
(m_\mu=\frac{i}{2}\tr (K(z,w)\,\sigma_\mu),\;\mu=0,\ldots,3)\\
\end{aligned}
\end{equation}
and, for the two eigenvectors $v_{\pm}$, with eigenvalue $\lambda_{\pm}$,
we have    
\begin{equation}
v_{\pm}=\frac{\textstyle 1}{\textstyle \sqrt{2m(m\mp m_3)}}
       \left(\begin{matrix}
              m_1-i m_2\\ 
              \pm m-m_3\\
        \end{matrix} 
          \right),\quad\lambda_\pm=m_0\pm m,\quad m=\sqrt{m_1^2+m_2^2+m_3^2}\, .
\end{equation}
The curvatures $F$ of $v_+$ is given by\footnote{The curvature of $v_-$ is minus that of $v_+$.}
\begin{equation}
F=\frac{i}{4}\epsilon_{ijk} \frac{m_i dm_j dm_k}{m^3}
\label{curvaturexpression}
\end{equation}
which we note vanishes for the bipartite case where $m_3=0$.
  
Note that in this non-bipartite case where $m_3\not=0$,
the points $p, \bar p\in T_2$ are now no 
longer excluded  since $\lambda_+$ is positive there;
thus the bundle $E^+$ now extends over all of $T_2$. 
Let $c_1(E^+)$ denote the value of  the first Chern class on $E^+$  so that
(some authors define $c_1(E^+)$ with the opposite sign to ours)
\begin{equation}
c_1(E^+)=\int_{T^2}\frac {i F}{2\pi} 
\end{equation}   
In order to conveniently display the values of the edge weights  we shall denote 
$c_1(E^+)$ by $c_1(a,b,c,d,D_1,D_2)$, in an obvious notation. Some selected  results for 
$c_1(a,b,c,d,D_1,D_2)$ are that 

\begin{equation}
\begin{aligned}
         \  c_1(1,1,1,1,1,1)=-1 & \\
             c_1(1.1,1.2,1.3,1,1,1)=-1 &\quad \text{(topological invariance)}\\
\end{aligned}
\end{equation}
other non-zero values are easily calculated. 
The value of  $c_1(a,b,c,d,D_1,D_2)$  is stable under  small changes in the 
edge weights; and when $c_1(E^+)\not=0$ it provides a certain topological
stability to the dimer configuration.   

Note that changing the Kasteleyn
orientation reverses  the sign of the diagonal terms in $K(z,w)$ and changes 
the sign of $c_1(a,b,c,d,D_1,D_2)$. A non-zero Chern number
therefore distinguishes between clockwise odd and
anti-clockwise odd Kasteleyn orientation.

We already know that when the graph is bipartite e.g., when $D_1=D_2=0$
the curvature $F$ and, hence the Chern number, both vanish.  However
the Chern number can also vanish in the non-bipartite case when the
edge weights are such that $m_1-i m_2\neq 0$ for any point on the unit
torus $T^2$, i.e. the edge weights specify a point off the associated
bipartite amoebae.

For suppose that 
\begin{equation}
c>a+b+d
\end{equation}
so that one is off the amoeba of $\hat P(z,w)$, then even when $D_1,D_2\not=0$,
one has 
$c_1(a,b,c,d,D_1,D_2)=0$. 
For example, setting $c=3.2$ and all other weights to unity yields 
\begin{equation}
c_1(3.2,1,1,1,1,1)=0
\end{equation}

In fact this is a natural result as the line bundle $E^+$ has a global non vanishing section $s$  given by 
\begin{equation}
\begin{aligned}
s&= \left<v, v\right>,\quad  v=\left(\begin{matrix}
              m_1-i m_2\\ 
              m-m_3\\
        \end{matrix} 
          \right)\\
&=2m(m-m_3)\\
\end{aligned}
\end{equation}
so that $s$ cannot vanish, since  
\begin{equation}
                 s=0\Rightarrow \begin{cases} m=m_3 & \text{i.e. $\hat P(z,w)=0$ , impossible off the amoeba} \\
                                              m=0   & \text{i.e. $m_3=\hat P(z,w)=0$, also impossible} \\
                                        \end{cases}  
\end{equation} 

In general the curvature $F$ will be non-zero for non-bipartite graphs
that have a subgraph on the bipartite amoeba. However, remember
from (\ref{curvaturexpression}) that 
\begin{equation}
m_3=0\Rightarrow F=0
\end{equation}
but   
\begin{equation}
m_3=i(D_1+D_2)(z-\bar z)
\end{equation}
so that $F=0$ when $D_1=-D_2$;  the connection $A=v^*_+ dv_+$ is then flat.
This connection has non-trivial holonomy and so is not trivial: indeed if
we choose the curve $C$ we had above, defined
by  $\theta=\theta_0$, $\theta_0$ constant, then with $c=a=b=d=1$   
\begin{equation}
\exp\left[\int_C A\right]=\left\{
                                  \begin{matrix} -\frac{1}{2}&\text{if $\theta_0\in\left(-\frac{\pi}{2},\frac{\pi}{2}\right)$}\\
                                                             &\\
                                                  \quad 0          & \text{otherwise}\\
                                   \end{matrix}
                           \right.
                      \end{equation}
and so one has   non-trivial holonomy as well as flatness.    

Reversing  the sign of one of $D_1$ and $D_2$
in (\ref{nonbipartitetwosite}) introduces vortices and
makes the curvature $F$ of (\ref{curvaturexpression}) zero.

However, reversing both $D_1$ and $D_2$ changes the sign of $F$.
It reverses the Kasteleyn orientation of the graph but the partition
function on the torus is unaffected. Hence the curvature, and Chern numbers are
sensitive to the Kasteleyn orientation, though the partition function is not.

These results are naturally interpreted via the real and complex K-theory of $T^2$---cf. the appendix for more details.
The K-theory statements for $T^2$ say that
\begin{equation}
  \begin{aligned}
  \widetilde K(T^2)&={\bf Z}\\
  \widetilde KO(T^2)&={\bf Z_2}\oplus {\bf Z_2}\oplus {\bf Z_2}\\
  \end{aligned}
\end{equation}
and this allows for both integer and ${\bf Z_2}$ invariants which is as one wants.
Hence flat connections with non-trivial holonomy are not  restricted to bipartite graphs; 
note that this example requires  that one of the edge weights is negative so that, as 
discussed in the next section, a {\it vortex} is present.

\section{Non Harnack  amoebae: singularities and area shrinking}
\label{SecNonHarnackAmoebae}
Let us assume,  for the moment, that our graphs $\Gamma$ are bipartite. 
If a plaquette of  $\Gamma$ contains one or more  {\it negative} edge weights, this can be compensated for by changing an arrow direction,  then the 
clockwise odd rule  will be violated for the plaquettes on
either side of this link: this can be interpreted as the presence
of vortices on these plaquettes.  The negative weight assignment
means that the curve ${\cal C}$ is non-Harnack. 

This can be detected in two equivalent ways:   
\begin{enumerate}[(i)]
\item  The amoeba map 
\begin{equation}
Ln:{\cal C}\longrightarrow {\bf R}^2 , \quad \text{(complex double fold  or pinch)}
\end{equation}
fails to be $2$ to $1$ for all points on the amoeba.
\item The amoeba satisfies 
\begin{equation}
{\rm Area\,}({\cal A(C)}) <  \pi^2{\rm Area\,}(\Delta) \quad \text{(Area shrinking)} 
\end{equation}
where $\Delta$ is the Newton polygon of ${\cal C}$.
\end{enumerate}
We shall now give some concrete examples of these phenomena.

\example{A complex double fold}
For our first example we take the graph shown in figure  \ref{vortexchainfig} below. Now, moving $z$ and $w$ off the unit torus and setting all edge weights
to unity, except for $J_{1z}$, which is set equal to $J$; the
resulting $K(z,w)$ is given by 
\begin{figure}
\begin{center}
\includegraphics[height=35mm]{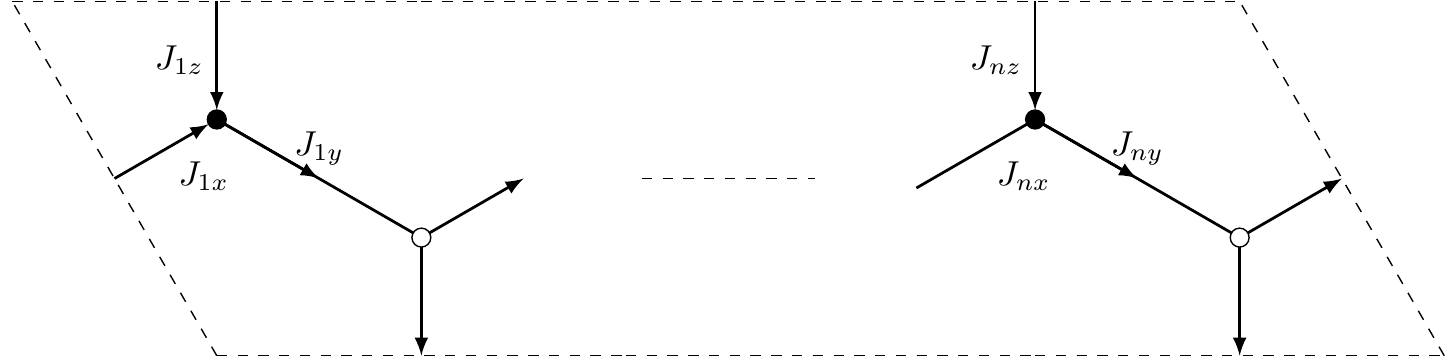}
\caption{A hexagonal chain of length $n$}
\end{center}
\label{vortexchainfig}
\end{figure}

\begin{equation}
K(z,w)=\left(\begin{matrix} 0&\hat A(z,w)\\
                       -\hat A^T(\bar z,\bar w)&0\\
                 \end{matrix}\right)_{2 n\times 2n}
\end{equation}
 and 
\begin{equation}
\hat A(z,w)=\left(\begin{matrix}
                 1-J w&0&\cdots& 0&-\frac{1}{z}\\
                       -1&1-w&0&\cdots&0\\
                        0&-1&1-w&\cdots&0\\
        \vdots&\vdots&\vdots&\ddots&\vdots\\
                         0&\cdots&0&-1&1-w\\
              \end{matrix}
\right)_{n\times n}
\end{equation} 
which yields 
\begin{equation}
\hat P(z,w)=(1 - J w) (1 - w)^{n-1} - \frac{1}{z},\;n=2,\ldots\quad .
\label{hexchain}
\end{equation}
We are interested in what happens when the edge weight $J<0$.
We do not consider  the case $n=1$ as this gives the hexagonal graph of
figure \ref{twoexamples} and does not yield a vortex when $J<0$: changing 
the sign of $J$ in this case gives an alternative Kasteleyn orientation of the
graph; the orientation is still clockwise odd  and ${\cal C}$ is still Harnack.

However when $n\ge2$ and $J<0$  we get two adjacent vertical columns of vortex
filled plaquettes followed by $n-2$ non-vortex columns in a periodic structure and we shall find that the $2$ to $1$ property of the
map ${\rm Ln}:{\cal C}\longrightarrow {\bf R}^2$  fails. 

Now we set $J=-1$ and turn first to the case $n=2$, in which case all
plaquettes on the lattice contain a vortex.  We have
\begin{equation}
\hat P(z,w)=1-w^2- \frac{1}{z}
\end{equation} 
and it is easy to check directly---or by comparison with the
amoeba of $\hat P(z,w)=1-w-1/z$---that ${\rm Ln}$ is now $4$ to $1$ 
{\it everywhere} and $2$ to $1$ nowhere. One can also easily check that
the area of the amoeba for $J=1$ is $\pi^2$, as required by
the Harnack condition, while for $J=-1$ it is $\pi^2/2$.

\begin{figure}
\includegraphics[width=75mm]{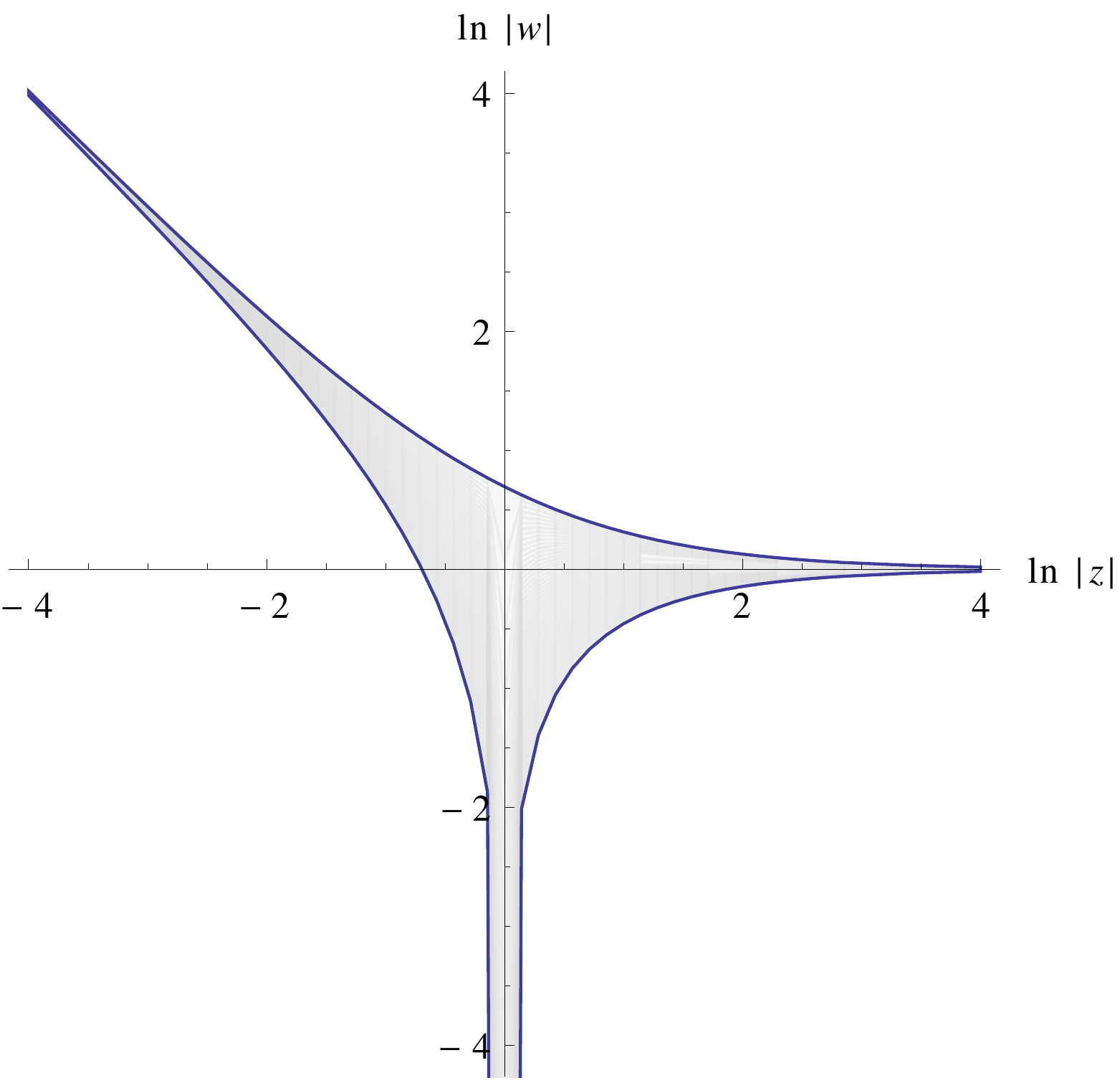}
 \hfill 
\raisebox{12mm}{\includegraphics[scale=0.55]{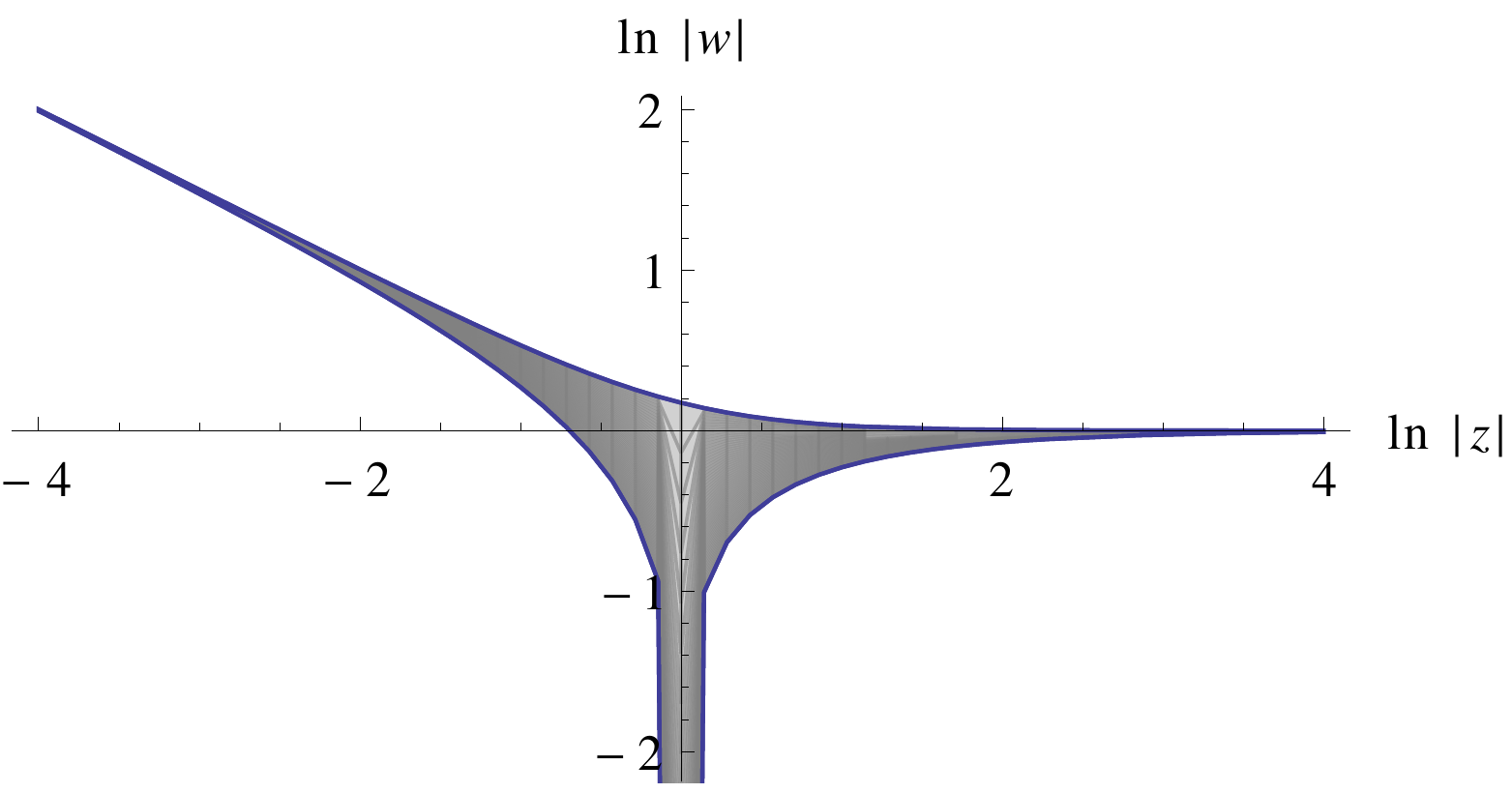}}
\caption{The amoebae for $\hat P(z,w)=1-w-1/z$ and $\hat P(z,w)=1-w^2-1/z$.  In
  the first figure ${\rm Ln}$ is $2$ to $1$ everywhere while in the
  second figure it is $4$ to $1$ everywhere.}
\label{comparisonfigures}
\end{figure}

We show the amoeba---together with that of $\hat P(z,w)=1-w-1/z$---in figure \ref{comparisonfigures}; 
the dark shading denotes the $4$ to $1$ region for ${\rm Ln}$. 

For the remaining values,  $n\ge3$,  the amoeba ${\cal A(C)}$ has both
a $2$ to $1$ region and a $4$ to $1$  region; the two regions being
separated by a complex double folding
(Mikhalkin  \cite{Mikhalkinsingularities}). We proceed to find the
singularity of   ${\rm Ln}$.

Quite generally a  singularity occurs when the  Jacobian ${\rm J(Ln})$
fails to have maximal rank everywhere on  ${\cal A(C)}$.
For the amoeba  we have 
\begin{equation}
\begin{aligned}
&{\rm Ln}:{\cal C}\longrightarrow {\bf R}^2\\
           &\left(f(w),w\right)\longmapsto 
\left(\ln\left\vert f(w)\right\vert,\ln\vert w\vert \right)\\
\end{aligned}
\quad\hbox{ where }\quad f(w)=\frac{1}{(1-J w)(1-w)^{n-1}}
\end{equation}
and our singularity condition is therefore
\begin{equation}
\det({\rm J(Ln)})=0
\end{equation}
where, if $w=\rho {\rm e}^{i\phi}$, one has   
\begin{equation}
{\rm J(Ln)}=\left(
               \begin{matrix}
                    \partial_\rho \ln\vert f\vert &\partial_\rho \ln\vert w\vert\\
                        \partial_\phi \ln\vert f\vert& \partial_\phi \ln\vert w \vert\\
                \end{matrix}
               \right)\, .
\end{equation}
We see then that  the amoeba is singular where  
\begin{equation}
\partial_\phi \ln\vert f\vert=0\, .
\end{equation}
Writing $\rho={\rm e}^y$ we obtain 
\begin{equation}
\frac{{\rm e}^y \sin(\phi) (J+J {\rm e}^{2y}+n-1-2 J {\rm e}^y \cos(\phi) n+J^2 {\rm e}^{2y} n-J^2 {\rm e}^{2y})}{
(-1+ 2 {\rm e}^y \cos(\phi)(J+1)-{\rm e}^{2y}(J^2+1+4 J \cos^2(\phi))+2 {\rm e}^{3y} \cos(\phi) (J^2+J)-{\rm e}^{4y} J^2)}=0
\end{equation}
and on setting $J=-1$ the solutions are 
\begin{equation}
\begin{aligned}
             \sin(\phi)&=0\\
          \cos(\phi)&=\frac{(2-n)\cosh(y)}{ n}\\
\end{aligned}
\end{equation}
meaning that  ${\rm Ln}$ is singular when
\begin{equation}
\begin{aligned}
            \phi &=0,\pi\\
            \phi &=\phi_*(y)\quad\text{where } \phi_*(y)=\arccos\left(\frac{(2-n)\cosh(y)}{ n}\right)\\
\end{aligned}
\end{equation}

\begin{figure}
\begin{center}
\includegraphics[scale=0.8]{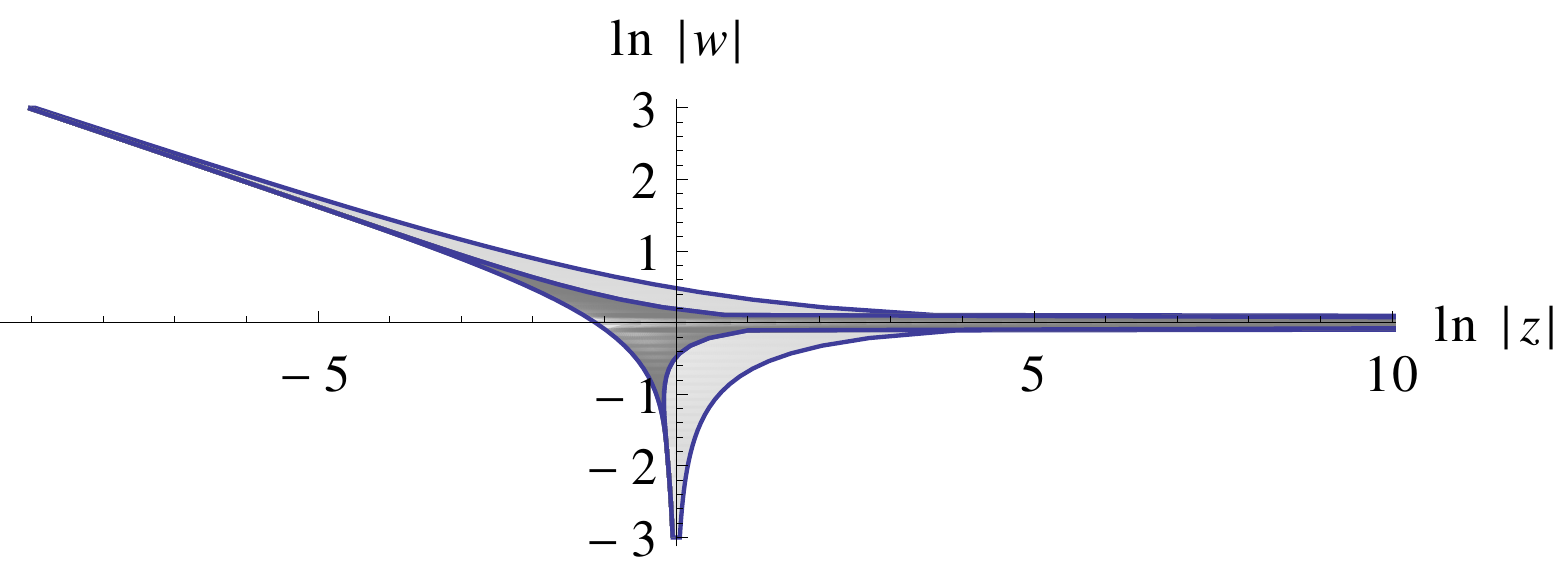}
\caption{A vortex amoeba with $\hat P(z,w)=(1 - J w) (1 - w)^{n-1} - 1/z$, ($J=-1,\,n=3$). The light and dark regions  are the areas where  ${\rm Ln}$ is $2$ to $1$ and  $4$ to $1$ respectively. The rightmost curve corresponds to $(z,w)=({\rm e}^{x},{\rm e}^y)$, while the interior curve is $(z,w)=({\rm e}^{x},{\rm e}^{y+i\pi})$ and this latter curve intersects the leftmost curve $(z,w)=({\rm e}^{x},{\rm e}^{y+\phi_*(y)})$ at finite
  values of $x$ and $y$, so that the lower boundary of the amoeba is composed of the finite segment of $\phi=\phi_*(y)$ and the
  unbounded segments of $\phi=\pi$.}
\end{center}
\label{complexdoublefold}
\end{figure}

We exhibit   an example of  the resulting amoeba in figure \ref{complexdoublefold}: 
in the dark region ${\rm Ln}$ is $4$ to $1$ and in the light region ${\rm Ln}$ is $2$ to $1$.
The  $4$ to $1$ region contains the amoeba origin $(0,0)$ and consequently intersects with the unit torus $T^2$. This means that $\hat P(z,w)$ vanishes at the $4$ points
constituting ${\rm Ln}^{-1}(0,0)$: these form $2$ complex conjugate pairs which we denote by $p,\bar p$ and $q,\bar q$. The connection $A$ on $\det E^+$  is now defined over the $4$-punctured torus $T^2_{p,\bar p,q,\bar q}$ and $A$ has the non-trivial holonomy
\begin{equation}
\exp\left[\int_C A\right]
\end{equation}
e.g., when $C$ encircles one of these $4$ points. The K-theory statement has now  enlarged:  one has   
\begin{equation}
\widetilde KO(T^2_{p,\bar p,q,\bar q})={\bf Z_2}\oplus{\bf Z_2}\oplus{\bf Z_2}\oplus{\bf Z_2}\oplus{\bf Z_2}
\end{equation}
reflecting the two extra zeroes.

It is interesting, from a physical point of view to analyse this example
in more detail. If we consider the vortex full lattice corresponding to the
hexagonal tiling of figure \ref{twoexamples}:  i.e. figure \ref{vortexchainfig}
with $n=2$,  weights from figure \ref{twoexamples} but 
$J_{2z}=-a$. We then obtain the polynomial
\begin{equation}
\hat P(z,w)=c^2-a^2 w^2-b^2/z\, ,
\end{equation}
where $z$ and $w$ are on the unit torus.
The partition function in this case can be analysed in detail using the
techniques of \cite{Nash_OConnor_jphysa:2009,Nash:1995ba,Nash:1996kn}.

For completeness let us continue to use the same weights, and  summarise the result in the vortex free case 
corresponding to $n=2$  in figure \ref{vortexchainfig}. For  $\hat P(z,w)$ we have 
$\hat P=(c-a w)^2-b^2/z$; which has zeros at $(z,w)=({\rm e}^{i \Theta/2},{\rm e}^{i \Phi})$ and $(\bar{z},\bar{w})$. 
There are $2NM$ dimers and, in the thermodynamic limit, the logarithm of the bulk partition function
per dimer,
$W=\frac{\ln Z}{2NM}$, is given by
\begin{equation}
W(a,b,c)=\ln c+\frac{\Theta}{\pi}\ln(b/c)
+\frac{1}{2\pi i}({\it li}_2(\frac{a}{c}{\rm e}^{i\Theta})
-{\it li}_2(\frac{a}{c}{\rm e}^{-i\Theta}))
\label{W_dilog}
\end{equation}
with 
\begin{eqnarray}
&&\sin(\Theta)=\frac{b}{2r}\ ,\quad \sin(\Phi)=\frac{a}{2r}\quad {\rm and}\\
&&\kern -24pt r=\frac{a b c}{\sqrt{(a+b+c)(-c+a+b)(c-a+b)(c+a-b)\, }}\, .
\end{eqnarray}

One finds
\begin{equation}
  \lim_{N,M\rightarrow\infty}\frac{Z(N,M)}{{\rm e}^{2NM W_{Vortex}(a,b,c)}}=Z_{Dirac}(\tau,\theta,\phi)
  =\frac{1}{2}\sum_{u,v=0}^{1/2}
\left\vert\frac{\theta[{}^{\theta+u}_{\phi+v}](0\vert\tau)}{\eta(\tau)}\right\vert \, ,
\end{equation}
with
\begin{equation}
  \tau=\frac{2N b}{M a}{\rm e}^{i(\Theta+\Phi)}\, .
\end{equation}
and $\theta[{}^{\theta+u}_{\phi+v}](0\vert\tau_v)$ and ${\eta(\tau_v)}$
are the Jacobi $\theta$-function and Dedekind $\eta$-function respectively. 
This is the partition function for a Dirac Fermion propagating on the
continuum torus with modular parameter $\tau$ and   a flat connection,  but with holonomies ${\rm e}^{2\pi i \theta}$ and
${\rm e}^{2\pi i \phi}$ round the cycles of the torus.

The result for the vortex case can be obtained rather simply from those
of the one tile example of \cite{Nash_OConnor_jphysa:2009} with polynomial $\hat P=c-a w -b/z$. The bulk free energy is given by
\begin{equation}
W_{vortex}(a,b,c)=\frac{1}{2}W(a^2,b^2,c^2)\, .
\end{equation}
There are now four zeros which come in complex conjugate pairs. These occur at
$(z,w)=({\rm e}^{i\Theta_v},{\rm e}^{i\Phi_v})$
and $(z,w)=({\rm e}^{i(\Theta_v)},{\rm e}^{i(\Phi_v+\pi)})$,
together with their complex conjugates, where $\Theta_v$ and $\Phi_v$
are obtained from $\Theta$ and $\Phi$ by sending $a$,$b$ and $c$ to
$a^2$, $b^2$ and $c^2$ respectively.

Expanding around the zeros of the polynomial one can easily
establish that for large $M$ and $N$ we have 
\begin{eqnarray}
\lim_{N,M\rightarrow\infty}\frac{Z_{Vortex}(N,M)}{{\rm e}^{2NM W_{Vortex}(a,b,c)}}
=\frac{1}{2}\sum_{u,v=0}^{1/2}
\left\vert\frac{\theta[{}^{\theta+u}_{\phi+v}](0\vert\tau_v)}{\eta(\tau_v)}\right\vert^2 \, .
\label{Z_vortex}
\end{eqnarray}
where 
\begin{equation}
\tau_{v}=\frac{N b^2}{M a^2}{\rm e}^{i(\Theta_v+\Phi_v)} \ .
\end{equation}

In general the holonomies $\theta$ and $\phi$  depend on the details of how the system is scaled to the continuum limit. For $a=b=c=1$ in \cite{Nash_OConnor_jphysa:2009}
we found that these holonomies depend  on the conjugacy class of
$L$ and $M$ ${\rm mod}$ $6$ with $(\theta,\phi)=(\frac{1}{2}-\frac{q}{6},\frac{1}{2}+\frac{p}{6})$ for $(L,M)=(q\; {\rm mod}\; 6,p\; {\rm mod}\; 6)$. When
$(\theta,\phi)=(0,0)$ then the term $u=v=0$ in (\ref{Z_vortex}) is zero, 
the expression is modular invariant and the system has central charge $c=2$; as
can be read off from the decrease  of the finite size effects in the cylinder limit.

In summary: The leading finite size correction to the vortex free
partition function is given by the continuum limit of a free Dirac 
Fermion.  In contrast for the vortex full configuration described
above, the finite size corrections corresponds to two Dirac Fermions;
however, the partition function is not a free sum over spin
structures, rather the spin structures are constrained to be equal, so
these form a rather natural Fermion doublet.

\example{A pinch}
Consider the graph of figure  \ref{twositebipartite} with
\begin{equation}
\hat P(z,w)=c-b\, z- \frac{a}{w}- \frac{d}{w z}
\label{Pfig_twositebipartite}
\end{equation}
This gives a Harnack curve ${\cal C}$
and ${\rm Ln}:{\cal C}\longrightarrow   {\bf R}^2$ is $2$ to $1$ everywhere.
However, if some edge weights are negative 
then the amoeba ${\cal A}({\cal C})={\rm Ln}({\cal C})$
can develop a {\it pinch} singularity at some point $p\in {\cal A}$:
at this point  ${\rm Ln}^{-1}(p)$ is no longer even discrete,
but continuous, as we shall now discover.

\begin{figure}
\begin{center}
\includegraphics[height=100mm]{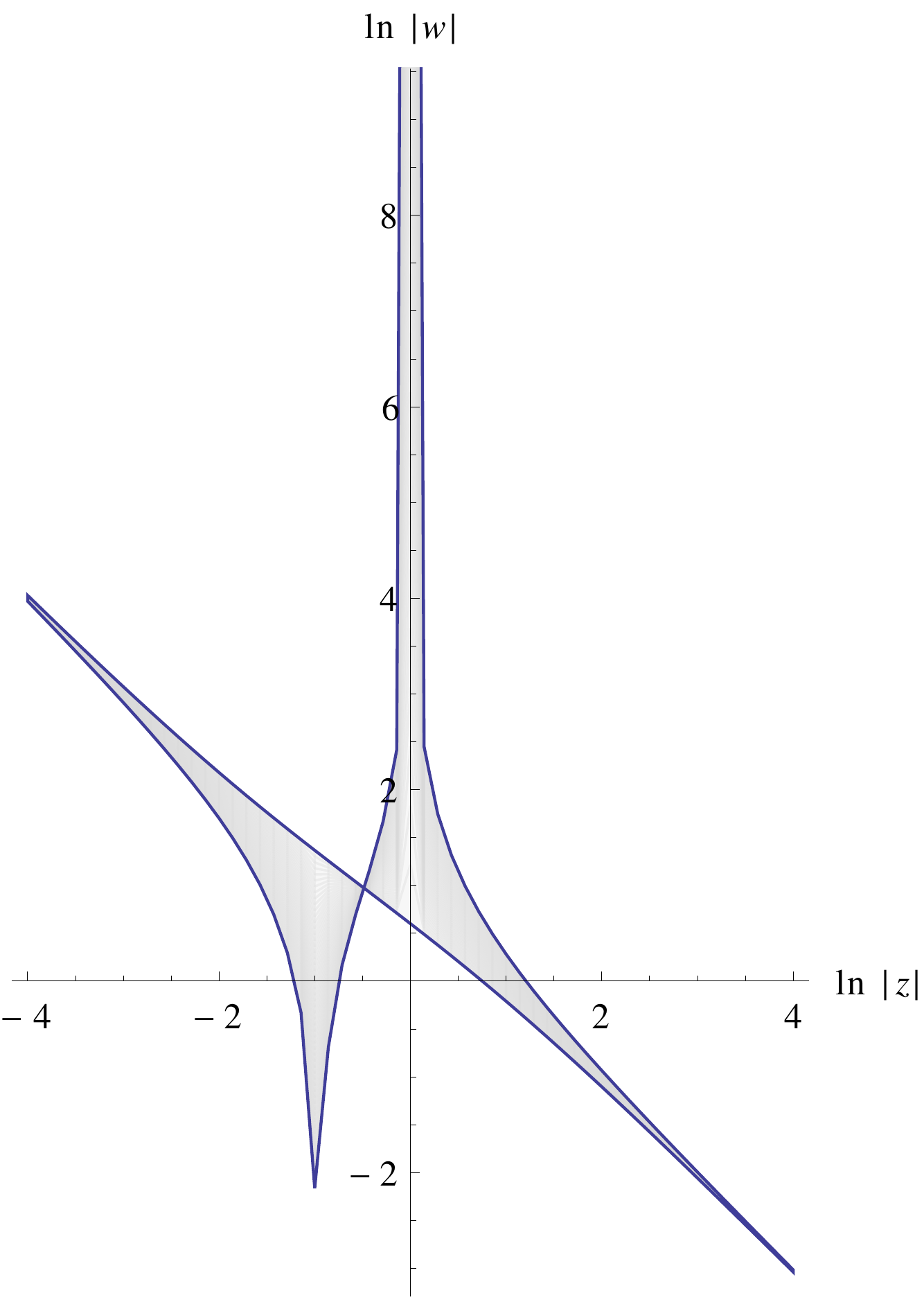}
\caption{A non-Harnack amoeba with a pinch for  $\hat P(z,w)=1- z- \frac{\textstyle a}{\textstyle w}- \frac{\textstyle 1}{\textstyle w z}$ and  $a=-\sqrt{7}$}
\end{center}
\label{pinchexample}
\end{figure}

Representing the curve ${\cal D}$ by solving $\hat P(z,w)$ in
(\ref{Pfig_twositebipartite}) for $w(z)$
the Jacobian condition 
\begin{equation}
\det({\rm J(Ln)})=0 \, ,
\end{equation}
with $z={\rm e}^{x+i\theta}$, $w={\rm e}^{y+i\phi}$ can be reduced to 
\begin{equation}
u(-\theta) \partial_\theta u(\theta) + u(\theta)\partial_\theta u(-\theta)=0,\qquad 
u(\theta) = \frac{a + d \,{\rm e}^{-x -i \theta}}{c - b \,{\rm e}^{x+ i \theta }} 
\end{equation}
This has the solutions $\theta=0,\pi$ which are the boundary of the amoeba; but 
it also has the  solution $x_p$ where 
\begin{equation}
x_p=-\frac{1}{2}\ln\left(-\frac{a b}{ c d} \right)\quad\hbox{ with }\quad y_p=\frac{1}{2}\ln(\frac{a^2}{c^2})\, .
\end{equation} 
This gives the pinch point $p$ which we write as  $p=(x_p,y_p)$  and ${\rm Ln}^{-1}(p)$ is a circle. If all edge weights are unity except $a$ then we find that
\begin{equation}
p=\left(-\frac{1}{2}\ln(-a),\ln(-a) \right)
\end{equation}
and we note the necessity for negative $a$.
We show a plot with the pinch in figure \ref{pinchexample}.

One can easily check that negative $a$ 
corresponds to a vortex full configuration on the lattice tiled with the fundamental tile of figure \ref{twositebipartite}.

\example{A shrunken area}
Next we come to an example where it is simple to demonstrate the area
shrinking imposed on a  non-Harnack curve. We  take the second graph on
figure \ref{twoexamples} for which 
\begin{equation}
\hat P(z,w) =a_1 c_2+a_2 c_1 + b_1 d_2 + b_2 d_1 - \frac{a_1 a_2}{z} - c_1 c_2 \,z - d_1 d_2 \,w - \frac{b_1 b_2}{w}\, .
\label{P_4sitesquare}
\end{equation} 
We have already considered this polynomial before and its amoeba is displayed 
in figure \ref{twoamoebae} for the case where 
\begin{equation}
\hat P(z,w)= 2D -z-\frac{1}{z}-w-\frac{1}{w},\quad D=2+2\cosh(t)
\end{equation}
This is a standard amoeba with
${\rm Area\,}({\cal A(C)})= \pi^2{\rm Area\,} (\Delta)\,$.

It turns out that the curve ${\cal C}$ is non-Harnack if $D<2$---an
impossibility in the present equation for $\hat P(z,w)$. However, if
$a_1=a_2^{-1}=-{\rm e}^{t_1}$,  $b_1=b_2^{-1}={\rm e}^{t_2}$ and all other 
weights are set to unity---so that we have a vortex
full lattice---then
\begin{equation}
  \hat P(z,w)= 2D -z-\frac{1}{z}-w-\frac{1}{w},\quad D=\cosh(t_2)-\cosh(t_1)
\end{equation}
and $D<2$ becomes accessible.

\begin{figure}
\begin{center}
\includegraphics[scale=0.45]{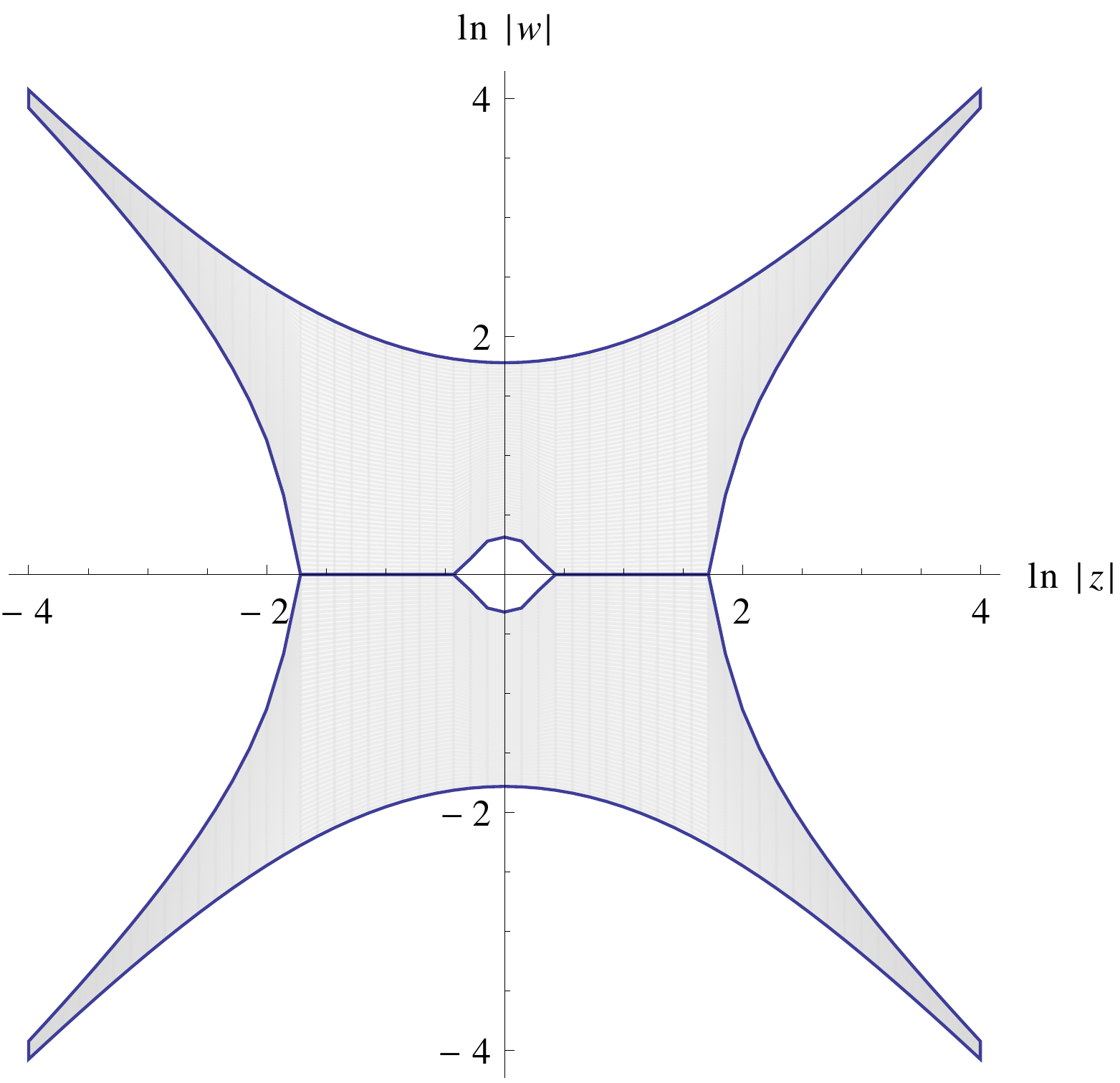}
\quad
\quad
\includegraphics[scale=0.45]{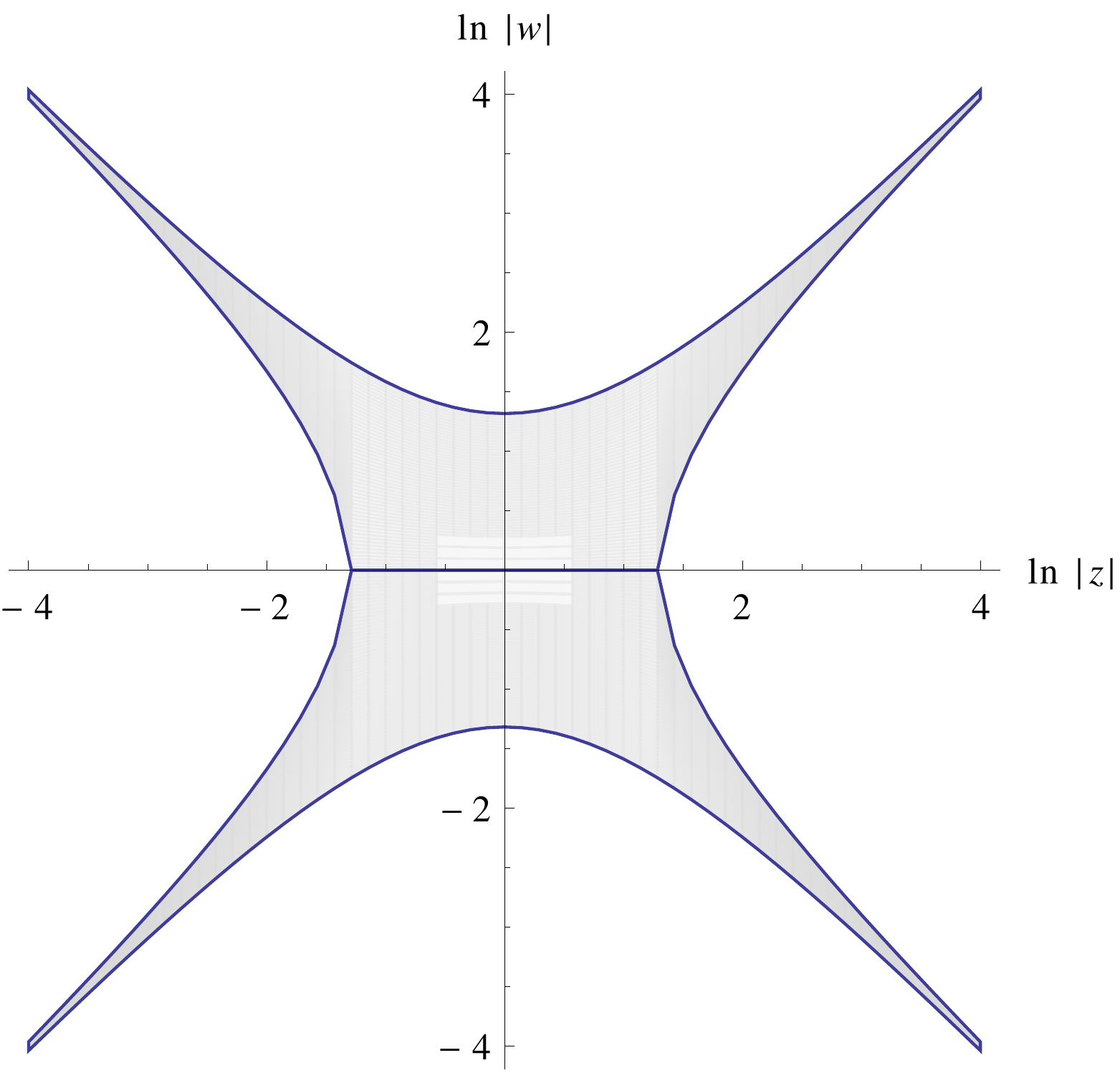}

\caption{The shrinking phenomenon displayed: both amoeba have  
$\hat P(z,w)=2 D-z-z^{-1}-w-w^{-1}$ but the larger one has $D> 2$ ($D=2.05$)  and is Harnack,  while the smaller one has $D<2$ ($D=1$) and is non-Harnack. }
\end{center}
\label{shrunkemarea}
\end{figure}

This means that ${\rm Ln}$ becomes singular but it 
also means that the second amoeba {\it shrinks}: denoting the two amoeba by  ${\cal A(C}_1)$ and ${\cal A(C}_2)$ respectively, 
one must have 
\begin{equation}
{\rm Area\,}({\cal A(C}_2)) < {\rm Area\,}({\cal A(C}_1)) =2\pi^2\, .
\end{equation}
In figure \ref{shrunkemarea} we show the ${\cal A(C)}$ for $D>2$ and $D<2$ and  the shrink is clearly manifest. Note that for $D\ge2$ the curve always has a compact oval and ${\cal A(C)}=2\pi^2$ while for $0<D<2$, we have ${\cal A(C)}<2\pi^2$. The case $D=0$ is rather special in that the
amoeba consists of the two lines $y=\pm x$  and ${\cal A(C)}=0$. 

We see that the topological type of ${\cal C}$ has degenerated
when $D<2$: it has lost a compact oval. For $D=2$  the curve ${\cal C}$
is still Harnack though there is a real node at the
origin $p=(0,0)$ of $ {\cal A(C)}$; but for $D<2$ there is a more
serious singularity and ${\rm Ln}^{-1}(p)$ is no 
longer discrete.

If we focus on the physical weights we see that introducing vortices,
as above, changes $D=\cosh(t_1)+\cosh(t_2)$ to $D=\cosh(t_1)-\cosh(t_2)$ 
so that for $t_2=0$, with $\cosh(t_1)< 3$, we are in the non-Harnack
case. Further---even in the presence of vortices---if $\cosh(t_2)=3$ we have $D=2$:  the degenerate Harnack case with no compact oval; but,  as $t_2$ is  increased still further, the curve is Harnack with a compact oval.
  
We shall now investigate the singularity  at the origin: choosing 
$z={\rm e}^{x+i\theta},\;w={\rm e}^{y+i\phi}$ 
we obtain 
\begin{equation}
\hat P(z,w)=2 D-2\cosh(x+i\theta)-2\cosh(y+i\phi)
\end{equation}
so that ${\cal C}$ is given by the equation
\begin{equation}
\begin{aligned}
D-\cosh(x+i\theta)-\cosh(y+i\phi)&=0\\ 
\Rightarrow y+i\phi&=\arccosh(D-\cosh(x+i\theta))\\
\end{aligned}
\end{equation}
Thus  $Ln$ is given by
\begin{equation}
\begin{aligned}
&{\rm Ln}: {\cal C}\longrightarrow {\bf R}^2\\
            &\qquad (z,w)\longmapsto (x, \Re(\arccosh(D-\cosh(x+i\theta)))\\
\end{aligned}
\end{equation}
and for the Jacobian  we have
\begin{equation}
J({\rm Ln})=\left(
               \begin{matrix}
                          1&0\\
                  \partial_x \Re(\arccosh(D-\cosh(x+i\theta)))&\partial_\theta \Re(\arccosh(D-\cosh(x+i\theta)))\\
                \end{matrix}
             \right)
\end{equation}
so  singularities occur when 
\begin{equation}
\begin{aligned}
\partial_\theta \Re(\arccosh(D-\cosh(x+i\theta)))&=0\\
   \Rightarrow \Re\left\{\frac{ -i\sinh(x+i\theta)}{\sqrt{(D-\cosh(x+i\theta))^2-1}} \right\}&=0\\
\end{aligned}
\end{equation}
Now suppose $x=0$, then  we have the condition 
\begin{equation}
\Re\left\{\frac{\sin(\theta)}{\sqrt{(D-\cos(\theta))^2-1}} \right\}=0
\end{equation}
and we see  the usual real  node solutions for   $\theta=0,\pi$ but, {\it also} solutions for those $\theta$  which satisfy
\begin{equation}
D-\cos(\theta)-1< 0
\end{equation}
which requires $D<2$. 

Now note that a point on the amoeba with $x=0$ has coordinates
\begin{equation}
(0,\Re(\arccosh(D-\cos(\theta))))=(0,0),\quad \text{when }D-\cos(\theta)< 1
\end{equation}
since $\arccosh(z)$  is pure imaginary when $z<1$. Hence ${\rm Ln}$ maps
all  these singular points to the amoeba origin $p$  
and ${\rm Ln}^{-1}(p)$ is no longer discrete when $D<2$ but consists of
the interval $[-\cos^{-1}(D-1),\cos^{-1}(D-1)]$. 
\section{Conclusions}
\label{Concludingsec}
We have found that the geometrical constructs of connection and
curvature can be very effective tools to analyse the structure of
dimer models, particularly on the torus with and without
punctures. One is led naturally to topological invariants including
holonomy, Chern classes as well as K-theory. Another effective tool,
to which we have frequent recourse, is the spectral curve ${\cal C}$:
an object which has dual life as a Harnack curve and the
characteristic polynomial of a bipartite dimer model. 
The amoeba
${\cal A(C)}$ of ${\cal C}$ also plays a central role: for example it
determines the phase diagram of the model and in the
presence of vortices it can even have singularities and facilitate the
uncovering of the rich Fermionic structure underlying dimer models.

Before finishing we wish to make an observation about Pfaffians and 
holonomy: in this work the Kasteleyn matrix, $K$, is a model for a
discrete Dirac operator and $\pfaff(K)$ is of central importance. Further when the vector bundle of positive eigenvalues of $i K$
has $\tr(F)=0$ the quantity $\tr(A)$---or just $A$ in the $U(1)$ case---is the 
simplest example of a Chern-Simons form and its exponentiated 
integral over $C$ is the holonomy invariant 
\begin{equation}
\exp\left[\int_C A\right]
\end{equation}
This expression is naturally a geometric invariant with values in ${\bf R/Z}$ just as in the higher 
dimensional Chern-Simons cases. 

The same two quantities turn up in studies  of global anomalies in the path integral for type II 
superstring theories with D-branes \cite{FreedWitten}. There the 
world sheet measure contains the  crucial product 
\begin{equation}
\pfaff(D)\exp\left[\int_{\partial \Sigma} A\right]
\end{equation}
with $\pfaff(D)$ the Pfaffian of the world sheet Dirac operator and 
$\partial \Sigma$ the boundary of the world sheet  $\Sigma$. 
There are also some intricate  discussions of holonomy and Pfaffians in 
\cite{Wittencondensedmatter}.
This parallel may repay further study. 

We have only just begun the study of dimer models with vortices.
As is evident from our study there is a rich structure to be
studied further here.

One could extend the study to models where lattice weights are
elements of a finite Abelian group, instead of just being real, or
complex, numbers.  Furthermore when dimer model partition functions
are realised as the Pfaffian of a Dirac operator one could further add
a gauge field in the form of holonomy elements linking the different
lattice sites.  This latter step would take us into the realm of
lattice gauge theory proper.

In summary, our current study has revealed that the presence of vortices
can alter the phase diagram of a dimer model, and change the
finite size corrections from a system with central charge, $c=1$ to one with
central charge $c=2$, corresponding to a Dirac doublet rather than the standard
vortex free case of a Dirac singlet.

Further properties and physical consequences of the presence of vortices
are discussed separately in \cite{inprep}.

\appendix
\section{K-theory}
\label{AppendixKtheory}
We present here some brief selected facts  on K-theory that are   more in
place in this appendix  than in the main body of the paper.

K-theory is a generalised cohomology theory of real, complex or quaternionic
vector bundles  over a base space $M$. We shall not consider quaternionic
vector bundles. K-theory places to the fore simplifications that occur when the
rank of the bundle is large enough compared to $\dim M$: the dimension of $M$. 

K-theory defines two rings $K(M)$ and $\widetilde K(M)$ arising from $\vbundm$ the set of all (isomorphism classes) of vector bundles over $M$ and these are related
by
\begin{equation}
K(M)= \widetilde K(M) \oplus {\bf Z}
\end{equation}
We shall be  concerned with $\widetilde K(M)$ which is called the reduced K-theory of $M$. The elements of $\widetilde K(M)$ are equivalence classes of vector bundles where, denoting an  equivalence class for a bundle $E$ by $[E]$,  two bundles $E$ 
and $F$ are equivalent (also called stably equivalent)  if the addition of a trivial bundle  to each of 
them renders them isomorphic: i.e. 
\begin{equation}
    E\oplus I^j\simeq F\oplus I^k
    \end{equation} 
we can then record this by writing $[E]\se [F]$.

In K-theory the  ring operations of sum and product are induced by direct sum and tensor product of bundles respectively---as required, multiplication is also distributive over addition. 

So far our bundles can be real or complex but now we shall distinguish between these two types.
We deal with the complex case first. Let $\hbox{\it Vect}_k(M,{\bf C})$ be the set of rank $k$ complex vector bundles over
$M$ and let $n=\dim M$ denote the real dimension of $M$. Then a key result  is: if $E_k\in \hbox{\it Vect}_k(M,{\bf C})$ and $p=[n/2]$---the smallest integer not greater  than $n/2$---then 
\begin{equation}
E_k\simeq F_p\oplus I^{k-p}
\end{equation}
for some rank $p$ bundle $F_p$.
One can check that this  means that
\begin{equation}
\hbox{\it Vect}_k(M,{\bf C})\simeq \widetilde K(M),\quad   \hbox{ for } k > n/2 
\end{equation}
So that making the rank $k$ of a  bundle $E$ larger than $n/2$ does not 
change the K-theory element $[E]\in \widetilde K(M)$; a  bundle with $k > n/2$ is said to be in the stable range.

Now we turn to real vector bundles---i.e. the set $\hbox{\it Vect}_k(M,{\bf R})$.
Here the key result is similar in character but the stable range is different.
One also needs some notation to distinguish  K-theory for real vector bundles from that for
complex vector bundles; we do this by writing\mod $\widetilde KO(M)$ for the real case
and $\widetilde K(M)$ for the complex case.
Now the key result is: if $E_k\in \hbox{\it Vect}_k(M,{\bf R})$---then 
\begin{equation}
E_k\simeq F_n\oplus I^{k-n},\quad\hbox{ when }k > n
\end{equation}
for some rank $n$ bundle $F_n$. This in turn yields the result that 
\begin{equation}
\hbox{\it Vect}_k(M,{\bf R})\simeq \widetilde KO(M),\quad   \hbox{ for } k > n 
\end{equation}
and the stable range for real vector bundles is therefore $k>n$.

Characteristic classes play an important role in  K-theory and, for complex vector  bundles, a prominent
role is played by the Chern character:  if, for simplicity, we specialise to the case the bundle $E$ has a  connection $A$ with curvature  $F_A$, then  the Chern character $ch\,(E)$ is defined by 
\begin{equation}
ch\,(E)=\tr\exp\left[\frac{i F_A}{2\pi}\right]
  \end{equation}
and it satisfies  
\begin{equation}
\begin{aligned}  
  ch\,(E\oplus F)&=ch\,(E)+ch\,(F)\\
  ch\,(E\otimes F)&=ch\,(E)\,ch\,(F)\\
\end{aligned}
\end{equation}
This in turn  means that the map
\begin{equation}
  \begin{aligned}
  ch:\;&\widetilde K(M)\longrightarrow \bigoplus_{i>0}H^{2i}(M;{\bf Q})\\
  {}  &[E]-[F]\longmapsto ch\,(E)-ch\,(F)\\
  \end{aligned}
\end{equation}
is a ring   homomorphism; while, for real vector bundles
 there is the ring homomorphism  
\begin{equation}
  \begin{aligned}
  ch:\;&\widetilde KO(M)\longrightarrow \bigoplus_{i>0} H^{4i}(M;{\bf Q})\\
  {}  &[E]-[F]\longmapsto ch\,(E)-ch\,(F)\\
  \end{aligned}
\end{equation}
However none of these maps  detects torsion in the K-theory.
Note that a complex vector bundle of rank $k$ has an underlying real
vector bundle $E_R$ of real rank $2k$.  

For the spheres $S^n$ one has the Bott periodicity results
\begin{equation}
  \begin{aligned}
    &\widetilde K(S^{n+2})=\widetilde K(S^n)\\
    &\begin{tabular}{|*{3}{c|}}
  \hline
  $n \mod2$ & 0  & 1  \\
  \hline
  \vrule height 12pt width 0cm 
  $\widetilde K(S^n)$ &{\bf Z}  &0  \\
\hline
    \end{tabular}
    \\
  \end{aligned}
  \qquad
  \begin{aligned}
    &\widetilde KO(S^{n+8})=\widetilde KO(S^n)\\
    &\begin{tabular}{|*{9}{c|}}
  \hline
  $n\mod 8$ & 0  & 1  & 2 & 3 & 4 & 5 & 6 & 7 \\
  \hline
  \vrule height 12pt width 0cm 
 $\widetilde KO(S^n)$ & {\bf Z}  & ${\bf Z}_2$ & ${\bf Z}_2$ & 0 & {\bf Z} & 0  & 0  & 0 \\
\hline
\end{tabular}
    \\
  \end{aligned}
  \end{equation}
and we notice  $\widetilde KO(S^n)$ contains torsion  even though $H^*(S^n;{\bf Z})$ is torsion free.

While, for general $M$, if $S\wedge M$, or $SM$ for short, denotes the reduced suspension of $M$ (which has the property that $SS^n\simeq S^{n+1}$); and one defines $\widetilde K^{-1}(M)$ by $\widetilde K^{-1}(M)=\widetilde K(SM)$ (and similarly for $\widetilde KO(M)$),  then one has 
$\widetilde K^{n+2}(M)=\widetilde K^n(M)$ and $\widetilde KO^{n+8}(M)=\widetilde KO^n(M)$.

When two spaces $X$ and $Y$ are joined at a point they are denoted by $X\vee Y$ and   one has
$\widetilde KO(X\vee Y)=\widetilde KO(X)\oplus \widetilde KO(Y)$ and similarly for $\widetilde K(X\vee Y)$. Thus for a
bouquet of circles one needs only $\widetilde KO(S^1)$ or $\widetilde K(S^1)$ as the case may be.

For Cartesian products $X\times Y$ one takes the space $X\wedge Y$---defined by  $X\wedge Y=(X\wedge Y/X\vee Y)$---and uses the fact that  
\begin{equation}
  \widetilde K^{-n}(X\times Y)= \widetilde K^{-n}(X\wedge Y)\oplus \widetilde K^{-n}(X\vee Y) 
\end{equation}  
and similarly for $\widetilde KO$.
It is now straightforward to calculate  the various K-theory rings that  we
require and, to this end, we would like to compare the real and complex K-theories of $S^2$ and $T^2$ for which we
find that 
\begin{equation}
  \begin{aligned}
    \widetilde K(S^2)&={\bf Z}\qquad \widetilde KO(S^2)={\bf Z_2}\\
    \widetilde K((T^2)&={\bf Z}\qquad \widetilde KO(T^2)={\bf Z_2}\oplus{\bf Z_2}\oplus{\bf Z_2}\\
  \end{aligned}
  \end{equation}
and we see that the complex  K-theories of $S^2$ and $T^2$ coincide but that the
real K-theories differ considerably.

For $S^2$ one also knows the appropriate
generators: if $H$ is isomorphic to the Hopf, or monopole line bundle,  over $S^2$ which has $c_1(H)=1$  then
$\widetilde K(S^2)$ has generator  $[H]-[I]$, whereas, if $H_R$ is the underlying real vector bundle of
rank $2$ to $H$ then  $\widetilde KO(S^2)$ has generator $[H_R]-[I^2]$. One can check explicitly that
$H_R\oplus H_R\simeq I^4$ so that  $[H_R]-[I^2]$ is of order $2$.

For $T^2$, if $f:T^2\longrightarrow S^2$ is a map of degree $1$,  then $f^*H$ is a  line bundle over
$T^2$ with $c_1(f^*H)=1$ and $[f^*H)]-[I]$ generates $\widetilde K(T^2)$; also the underlying real bundle
$f^*H_R$ will provide one of the generators of $\widetilde KO(T^2)$.

We have torsion in our holonomy calculations so we make recourse to
$\widetilde KO$ and observe that, for the punctured tori, which are
bouquets of circles, the above implies that
\begin{equation}
  \begin{aligned}
    \widetilde K(T^2_{p,\bar p})=0\\
    \widetilde K(T^2_{p,\bar p,q,\bar q})=0\\
\end{aligned}
\end{equation}
while for $\widetilde KO$ one has 
\begin{equation}
  \begin{aligned}
    \widetilde KO(T^2_{p,\bar p})&={\bf Z_2}\oplus {\bf Z_2}\oplus {\bf Z_2}\\
    \widetilde KO(T^2_{p,\bar p,q,\bar q})&={\bf Z_2}\oplus{\bf Z_2}\oplus{\bf Z_2}\oplus{\bf Z_2}\oplus{\bf Z_2}\, .\\
  \end{aligned}
  \end{equation}

The bundle $E^+_R$ has an Euler class $e(E^+_R)$, and Stieflel-Whitney classes $w_1(E^+_R)$ and $w_2(E^+_R)$:  these classes possess the properties 
$e(E^+_R)=c_1(E^+)$,  $w_2(E^+_R)=c_1(E^+) \mod 2 $  and $w_1(E^+_R)=0$  since $E^+_R$ is oriented. 

 In the case of non-bipartite graphs---cf. figure  \ref{twositenonbipartite} above---we found that
 the complex line bundle $E^+$ over $T^2$ had $c_1(E^+)=-1$, which is fine for $\widetilde K(T^2)$. Thus $e(E^+_R)=1$, $w_2(E^+_R)=1$ and $w_1(E_R^+)=0$. 
However to detect the ${\bf Z_2}$ holonomy
around the  two homology cycles, which turns up when the curvature vanishes, we should pass from $E^+$ to $E^+_R$ and use $\widetilde KO(T^2)$. 

For bipartite graphs $E^+$ one has $c_1(E^+)=0$, while
$e(E^+_R)=0$ and $w_2(E^+_R)=w_1(E^+_R)=0$.

\end{document}